\documentclass{article}

\usepackage{arxiv}
\usepackage{comment}
\usepackage[utf8]{inputenc} 
\usepackage[T1]{fontenc}    
\usepackage{hyperref}       
\usepackage{url}            
\usepackage{booktabs}       
\usepackage{amsfonts}       
\usepackage{nicefrac}       
\usepackage{microtype}      
\usepackage{lipsum}
\usepackage{amsmath}
\usepackage{graphicx}
\usepackage{enumitem}
\usepackage{caption}
\usepackage{subcaption}
\graphicspath{{figures/}}

\title{Bias-Resistant Social News Aggregator Based on Blockchain}

\author{
  Amir Ziashahabi \\
  Department of Computer Engineering\\
  Sharif University\\
  \texttt{ziashahabi@ce.sharif.edu} \\
   \And
 Mohammad Ali Maddah-Ali \\
  Department of Electrical Engineering\\
  Sharif University\\
    \texttt{maddah\_ali@sharif.edu} \\
    \And
 Abbas Heydarnoori \\
  Department of Computer Engineering\\
  Sharif University\\
    \texttt{heydarnoori@sharif.edu} \\
}

\begin{document}
\maketitle

\begin{abstract}
In today's world, social networks have become one of the primary sources for creation and propagation of news. Social news aggregators are one of the actors in this area in which users post news items and use positive or negative votes to indicate their preference toward a news item. News items will be ordered and displayed according to their aggregated votes. This approach suffers from several problems raging from being prone to the dominance of the majority to difficulty in discerning between correct and fake news, and lack of incentive for honest behaviors. In this paper, we propose a graph-based news aggregator in which instead of voting on the news items, users submit their votes on the relations between pairs of news items. More precisely, if a user believes two news items support each other, he will submit a positive vote on the link between the two items, and if he believes that two news items undermine each other, he will submit a negative vote on the corresponding link. This approach has several desirable features: (1) This will reduce the contribution of personal preferences on voting. One may disagree with two news items but still, find them supporting each other. (2) In the resulting graph, all news items related to one news item, act as evidence for that item, endorsing or disputing it. This approach helps the newsreaders to understand different aspects of a news item better. We also introduce an incentive layer that uses blockchain as a distributed transparent manager to encourages users to behave honestly and abstain from adversary behaviors. The incentive layer takes into account that users can have different viewpoints toward news, enabling users from a wide range of viewpoints to contribute to the network and benefit from its rewards. In addition, we introduce a protocol that enables us to prove fraud in computations of the incentive layer model on the blockchain. Ultimately, we will analyze the fraud proof protocol and examine our incentive layer on a wide range of synthesized datasets.
  \end{abstract}


  \section{Introduction}
In the past, a few newspapers and news agencies were in charge of generating and distributing news, making them the primary source of information for most people. With the advent of the web, individuals have become involved in the process, too, and their contributions have raised consistently to the extent that each day, more and more people receive the latest news from social networks. Social news aggregators are a kind of social network devoted to news in which users post news items and vote to promote them. As examples, Reddit and Hacker News can be mentioned. Social News Aggregators engage individuals in the process of collecting and popularizing the news and reflect their interests and comments.

  \subsection{Network Characteristics}
  \label{sec:problems}
Current social news aggregators are susceptible to several problems:

\begin{itemize}
	
	 \item Keeping users in filter bubble: Social networks, in general, try to satisfy their users by showing them the contents they are most likely to enjoy. So, users are always presented with their favorite side of a topic, and as a result, kept inside a filter bubble \cite{pariser2011filter}, which deprives them of seeing different aspects of a topic.

\item Dominance of the majority: Social news aggregators use the votes, which are cast on news items, as the main indicator of their importance and one of the main factors for prioritizing news items for presentation to the news readers. Therefore, if a group of united (and possibly adversary) users have a bias in specific topics, they can deviate the way others see items related to those topics through up-voting desired items and down-voting opposing items. Since the votes cast by other users do not have a specific pattern and are distributed over different news items and different viewpoints, the effect of united users can be quite significant even if they do not constitute the majority of users. Although this problem is mitigated by creating separate subsections wherein users with similar viewpoints are gathered, this approach makes it more difficult for users to pop the filter bubble since their exposure to other ideas is limited significantly. For the sake of simplicity, we refer to the dominant users as the majority in the rest of this paper.

\item Fake news: Fake news items were shared more than 37 million times during the 2016 United States presidential election \cite{allcott2017social}. Current social networks rely mostly on users' reports, artificial intelligence methods, and third-party fact-checking as the primary means of detecting fake news. So, the network runners are in charge of finding and deleting fake news. Indeed there is no built-in mechanism to help users, themselves, to discern truth from lies. 

\item Lack of incentives: current networks provide no incentive for users to act honestly, preventing them from reaching the full potential that active and honest users can bring to the network.

\item Centralized Management: the company managing the network decides on the way news items should be presented to users, which can be affected by politics and business.

\item Lack of transparency: The operators of the social network can modify, delete, or add votes on news items without a trace.

  \end{itemize}

With these problems in mind, the purpose of this study is to propose an alternate design for an SNA with the following characteristics:

\begin{itemize}
\item The voting mechanism is such that it reduces the effects of political views and personal preferences in voting. In current SNA platforms, the viewpoint of the users will directly affect their votes.

\item It allows us to observe and evaluate each news item in relation to other news items. It is in contract with the conventional approach that each news item is evaluated individually.

\item It is resistance to the dominance of the majority, and the platform reflects the minorities' viewpoints alongside the majority's.

\item It exposes an individual to a news item's context and the news items against their views. This will help them pop the filter bubble, besides giving them the opportunity to judge the veracity of news items better and detect fake news.

\item It does not allow a system administrator to eliminate the news items he deems undesirable, or change voting results as he wishes.

\item It operates in full transparency and every user is aware of the state of the network and is able to validate it.

\item It will discourage random or even adversary behaviors in voting and encourage users to express their honest opinion.

\end{itemize}

\subsection{Blockchain} \label{sec:blockchain}

Blockchain is the cornerstone of our proposed network. In this section we present an a brief introduction blockchain. Together with Bitcoin, a new data structure named blockchain was introduced in \cite{nakamoto2008bitcoin}, which allows the Bitcoin network to be run without trusting a central authority like a bank. In the Bitcoin network, the ledger is in the form of a sequence of blocks, chained one after another, where each block holds a set of transactions. These Blocks are produced and maintained by a set of volunteers called miners. Each block is generated by one of the miners, who wins in a puzzle-like competition. This block of transactions is double-checked by other miners to ensure it does not hold any invalid transactions. If the block is verified, it will be added to the blockchain. The winner receives some prizes, called Bitcoin, that can be spent later. 

Bitcoin is the first blockchain network that offers a trustless money transfer service that was previously only possible through centralized and trust-oriented systems, namely banks. The notion of removing the central authorities using blockchain can be expanded to a wide range of applications other than simple money transfer. The idea is that arbitrary contracts (applications) can be converted into computerized codes and have the blockchain, as a trustless decentralized judge, check the contract conditions and execute the terms accordingly. These computerized contracts are called smart contracts. Bitcoin was designed with the sole purpose of payment systems in mind, and its limited programming language fails to offer the expressiveness required for coding complex contracts. In \cite{buterin2014ethereum}, Buterin introduced Ethereum as the second generation blockchain, which benefits from a Turing complete programming language for creating smart contracts.  Every user can create and submit her own smart contract to the Ethereum network. After confirmation in the network, the smart contract cannot be changed, and the network guarantees that the logic is executed exactly as it is coded in the smart contract.

The remainder of the paper is structured as follows. In the next section, we provide a high-level picture of the proposed solution and the design challenges. In Section \ref{sec:solution-details}, the details of the proposed network are delineated, and Section \ref{sec:evaluation} contains the evaluation results of the mechanism proposed for the network. In Section \ref{sec:related-works} the related works are reviewed and Section \ref{sec:threats to validity} iterates over the threats to validity. The paper ends with some suggestions for future work and a conclusion.

\section{An Overview of the Proposed Solution}
In this section, we provide an overview of the proposed network to satisfy the characteristics mentioned in the previous section. The proposed solution consists of three folds. First, to achieve a more unbiased voting system that preserves the minorities' viewpoints and at the same time adds context to news items, we propose a new design that consists of a graph of news items connected by weighted edges. The edges and their weights are determined by users' votes and convey the relationship between news items. Second, in order to bring transparency to the network and eliminate the almighty administrators from the network, we utilize blockchain as the management layer of our proposed network. Finally, we introduce an incentive mechanism based on cryptoeconomics, which is empowered by the blockchain infrastructure, to motivate users to vote according to their honest views. Furthermore, we propose a layer-2 blockchain solution which, through which expensive computations of the incentive mechanism are done off-chain and their validity is enforced on-chain.

\subsection{News Graph}

The proposed approach is a graph-based social news aggregator in which users vote on the relations between news items instead of the items themselves. Based on how much two news items corroborate or undermine each other, users cast votes from $\{-1,0,1\}$, where 1 indicates that the two news items corroborate each other, -1 means that they oppose each other, and 0 means that the news items are neutrally related. Such a network creates an unfavorable setting for united users to undermine undesired news items. That is because the votes do not indicate the significance of a news item or how much people found it interesting; instead, they indicate its relationship with other items. In this situation, down-voting a relation does not decrease its importance; on the contrary, it makes it more important as it acts as a piece of undermining evidence for another item. Besides, the constructed network connects each news item to its related items providing the user with information from several aspects and sources, which makes it much easier for users to investigate the veracity of the news.

\subsubsection{A Toy Example}
In this section, we present a simple example of the expected news graph. Suppose there are six news items with titles: 
\begin{enumerate}
	\item Paris climate agreement entered into effect in 2016.
	\item US withdraws from the Paris climate agreement.
	\item US economy will rise in the future.
	\item Global warming has a negative impact on the economy.
	\item Paris climate agreement leads to job loss
	\item Global warming causes global sea level to rise.
\end{enumerate}
and two users:
\begin{enumerate}
	\item Alice is an environment activist
	\item Bob is a nationalist
\end{enumerate}

\begin{figure}
	\centering
	\includegraphics[width=0.8\linewidth]{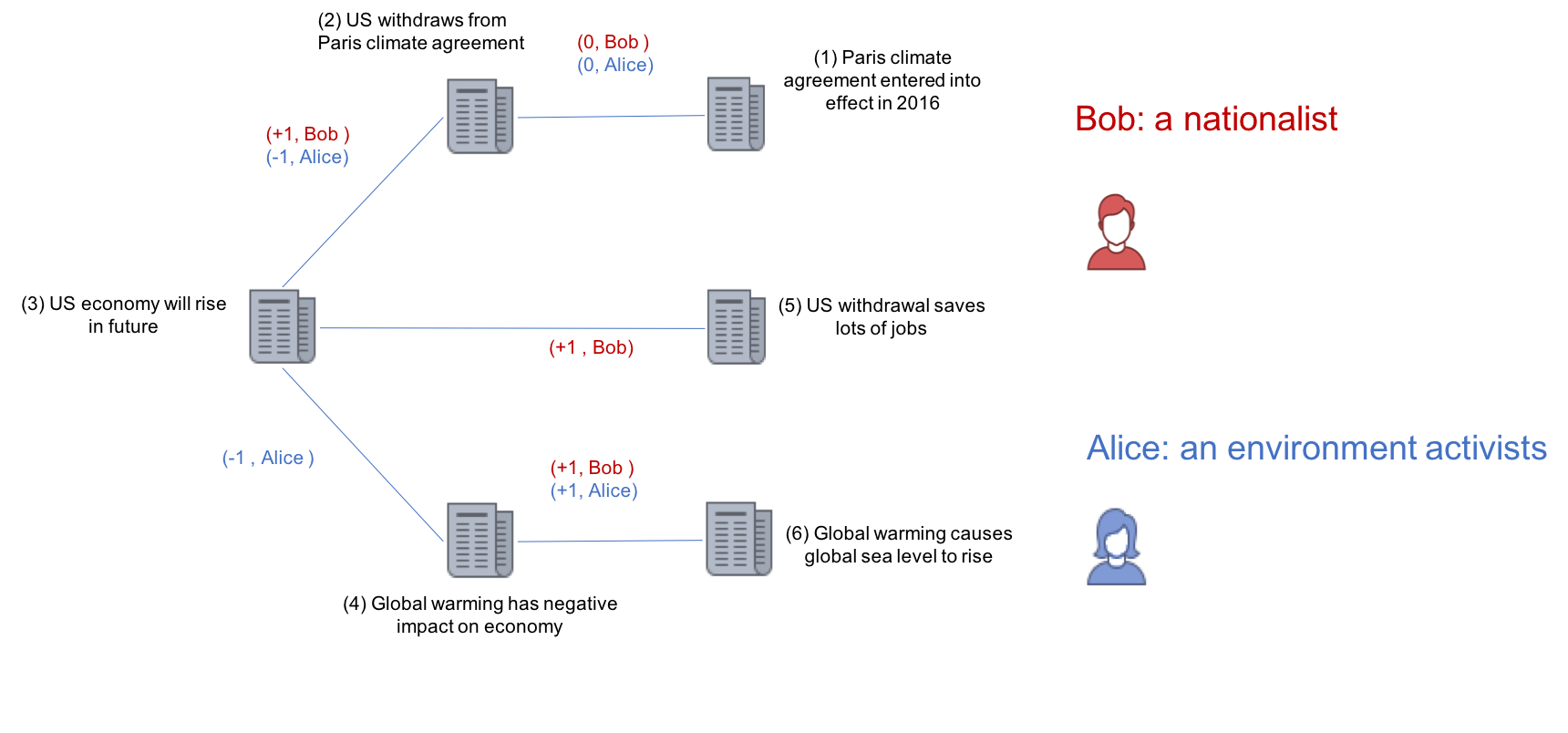}
	\caption{News items relations example.}
	\label{fig:fig1}
\end{figure}

Figure \ref{fig:fig1} shows the news items and the votes that users submitted.

Regarding the difference of opinions between Alice and Bob, several cases can occur in voting. In most cases, users vote on the edges that they are interested in without conflict. For example, they both agree that items 1 and 2 are related in a neutral way, or the rise of global sea level, as a result of global warming, can damage the economy (positive relation between items 4 and 6). However, there are cases where users might vote differently depending on their point of view. In the example, Alice believes that the US's withdrawal from the Paris agreement can exacerbate global warming, which will, in turn, have a negative impact on the economy in the long term (negative relation between items 3 and 4 as well as 2 and 3). On the other hand, Bob thinks that the US's withdrawal saves lots of jobs (positive relation between items 3 and 5), and it will strengthen the US's economy (positive relation between items 2 and 3). The difference in the user's viewpoints leads to a contradicting vote on the relation between items 2 and 3.

\subsection{Blockchain as the Manager}

Blockchain enables us to create applications that operate in a transparent and decentralized environment. In our use case, news items (in the form of hashes representing their addresses on a decentralized file platform like IPFS) as well as users' votes, are committed on the blockchain. Since a blockchain system is not managed by a single person but by a collective consensus mechanism, when a news item or a vote is recorded on the blockchain, it is locked into a semi-immutable state, which becomes harder and harder to change as time goes on.  

In addition, through using blockchain, news items are publicly available, and users can develop their desired tools to access a news item and its related items. This will give them the options to choose how and based on what algorithm the news items should be presented to them. Such construction will enable them to pop the filter bubble and explore news from different aspects.

\subsection{Incentive Mechanism}

In a blockchain network, there is no central authority to monitor and manage users, and they are allowed to act arbitrarily, and possibly maliciously. Therefore, it is necessary to devise mechanisms that drive people to act honestly and vote on news item relations the way they actually think. On the bright side, the cryptoeconomics, as one of the gifts of blockchain, provides effective ways to tackle this problem. When using blockchain, we are able to create our own currency as a token, a privilege that is not achievable using a centralized system. The reason is that centralized networks are incapable of guaranteeing that the tokens are only minted following specific laws, and as a result, in these networks, tokens can be minted whenever the managers decide. Blockchain has solved this problem, and token creation is one of the most popular applications of blockchain. We utilize this feature in combination with a collaborative filtering method to design an incentive mechanism for our network. In the resulting mechanism, users' behaviors are analyzed, and they are rewarded based on their behavior. This is done in a way that each user is incentivized to act honestly and promote the network's quality in order to gain more financial benefits (in terms of tokens).

\subsection{Challenges}
In this section, we discuss the challenges in the way of the design of the proposed network.

\subsubsection{Incentives for honest behavior}

One of the main characteristics of public blockchains is that they enable us to govern the network by taking advantage of cryptoeconomics protocols so that for every user, it is economically best if they follow the protocol and act honestly. The objective here is to determine whether a user is acting honestly or not and then reward them accordingly. But, how can we decide what kind of behavior is honest behavior? The main challenge is that votes are subjective, so we cannot determine what the correct votes are ( the votes that an honest user is expected to submit). A relation that is corroborating from one person's perspective can be undermining seen by another person. For example, consider the simple approach of using the majority of the votes on a relation as the indicator of the correct behavior. The problem with this solution is that it does not respect the difference in opinions of users and considers a specific behavior, which is manipulable by the majority, as the right behavior that should be expected from everyone.

The question here is to what degree users should have freedom in their voting behavior. For example, is it acceptable for a user to vote randomly? The answer is no, since this kind of voting adds no beneficial information to the network, and allowing this behavior paves the way for spammer and adversaries to fill the network with useless votes.

So, The protocol should limit the power of the majority to dictate a single right behavior, and voices with different views should be valued. On the other hand, the protocol should prevent people from injecting detrimental, effortless info into the network by voting randomly. The challenge here is how we can differentiate logical but different behavior from random voting.

\subsubsection{Computation Limit}
One of the challenges in the design of a blockchain-based network is the limited processing power of blockchain. In the initial blockchain platforms, like Bitcoin \cite{nakamoto2008bitcoin} and Ethereum \cite{buterin2014ethereum}, every transaction in every block is verified by each node (miner) in the network. The more computation-heavy blocks, the less motivation for miners to validate the blocks. Besides, bigger blocks take more time to propagate through the network, hence increase the chance of fork creation and network partitioning. Therefore, blockchain networks are designed in a way that the processing power of the network is kept constant and does not scale so that blocks propagate faster, and nodes with lower computation power can contribute to the network resulting in the increased security of the network. This limited processing power is not sufficient for the computation-intensive operations required in our incentive mechanism.

\section{The Proposed Solution in Details} \label{sec:solution-details}

In this section, the details of the proposed network will be delineated in three versions. We start with a trusted centralized network, and as the version increases, more complexity and details are added to the network.

\subsection{Version 1} \label{sec:version-1}
The focus of this version is on the details of the incentive mechanism. For the sake of simplicity, we assume that the network has a trusted centralized manager, which receives votes, analyzes users' behaviors, and rewards them accordingly. First, the news graph's formulation is presented, and then the incentive mechanism is explained.

\subsubsection{News Graph Formulation} \label{ssec:graph-fromulation}

We are given an undirected news graph $G_s = (\mathcal{N},\mathcal{E})$ and a set of users $\mathcal{U} = \{u_1, u_2, ..., u_{|U|}\}$. News items $\mathcal{N} = \{n_1,n_2,..., n_{|N|}\}$ are connected with edges $e(n_1,n_2) \in \mathcal{E}$ where $n_1,n_2 \in \mathcal{N}$. Users vote on the edges and we define votes matrix $R \in \mathbb{R}^{|\mathcal{U}| \times |\mathcal{E}|}$ as:
\begin{align*}
&\mathbf{r}_{i,j} \in \{-1,0,1\} &&\text{if user $i$ has voted on relation $j$}\\
&\mathbf{r}_{i,j} = * &&\text{otherwise}
\end{align*}
To have a smaller $\mathcal{E}$, and avoid $|\mathcal{E} | = |\mathcal{N}|^2$ the edges that have no votes are removed.

\subsubsection{Incentive Mechanism}

As mentioned before, subjectiveness is the property that entails the main challenge for designing the incentive mechanism, which requires the mechanism to value individual views while preventing random voting. If a user votes honestly, she is expected to roughly follow a discernible pattern in her behavior. The assumption that users' behaviors can have detectable patterns is the basis for the ubiquitous recommender systems. In this paper, we propose an incentive mechanism that takes advantage of this assumption to detect honest users from random voters. To put it in other words, the purpose of the proposed incentive mechanism is to detect whether a user has a specific pattern in her behavior, or she is voting randomly. If a user's behavior conforms to a pattern, we consider the user as having a consistent behavior. The question is, how can we determine whether the voting pattern of an individual has consistency or not?

To answer this question, we use a collaborative filtering approach. Collaborative filtering methods aim to detect user-item relations depending solely on the history of the users and their interaction with items. As a result, this approach eliminates the prohibitive need for explicit characteristics of users and items \cite{koren2009matrix}. This is an essential feature that enables us to judge users' behaviors solely based on the only available info, which is their voting behavior.

One of the categories of collaborative filtering is the latent factor models. Latent factor models try to model users and items using a number of latent factors \cite{koren2009matrix}. These factors are deduced from the user-item relations and might correspond to an interpretable or a non-interpretable characteristic. Now that the model can capture the characteristics of users and items, it will be able to predict what will be the view of a user toward an item. 

In our case, the items that we are interested in are the new items relations. With respect to the fact that latent factor models take into account the characteristics of users and relations, they fit very well into the requirement of our incentive mechanism. Following this approach, we do not care if the user's votes are consistent with the majority, but rather we try to detect users' characteristics and check whether they are consistent with the user's voting behavior.

Before we delve into the details, staking should be explained.

\subsubsection{Staking} \label{sssec:staking}

In order to incentivize honest behavior, we need a means of keeping users accountable for their actions. Staking is one of the prevalent techniques in the cryptocurrency environment used for this purpose. In this regard, we use an approach similar to the mechanism proposed in \cite{steemit}. According to this approach, in order to contribute to the network, each user should buy the network's cryptocurrency token and stake some of it. Stakes can be considered as a long term investment, and a stake withdrawal request by the owner will be fulfilled after several months. Using stakes as the weight of users in the model used by the incentive mechanism encourages honest behavior and discourages adversary behaviors. The more token a user stakes in the network the more benefits he gain from the increase of the token price, so is more likely to act honestly.  On the other hand, having stakes integrated into the model disincentivizes adversary behaviors by making the model more robust and increasing the cost of such behaviors. The reason is that in order to compete with the honest users, and be more determinant in the model's outcome, an adversary has to stake a significant amount of tokens; knowing that his success in damaging the network will lead to the plummet of network's token price, and consequently, by the time he is allowed to withdraw, his stakes will be worth much less. To limit the impact of each user in the model, users can stake up to 5 tokens.

\subsubsection{Detecting Consistent Behavior Using Collaborative Filtering}

As mentioned in Subsection \ref{ssec:graph-fromulation}, we represent the news graph as matrix $\mathcal{R}$, which contains the votes of users to news items relations. To analyze this matrix, we use matrix factorization as a latent factor model that is simple to validate and handles sparse, large-scale data well \cite{ricci2011recommendation}. The idea is that, analogous to prevalent user voting system's data, such as movie rating systems, we can presume that users and the relations between news items can be characterized using latent factors. So, if we find a good enough approximation of the latent features, people whose votes deviate from the approximation can be considered as having inconsistent and pattern-less behaviors.

By incorporating users' stakes into the matrix factorization problem, we formulate the problem as follows:

\begin{align} \label{eq:MF}
\begin{split}
&{\underset {\mathbf{\tilde R}}{\operatorname {arg\,min} }}\,\|\mathbf{R}-{\tilde {\mathbf{R}}}\|_{\rm {C}},\\
&\text{s.t. } \; \; \tilde {\mathbf{R}} = \mathbf{U} \times \mathbf{V}^T
\end{split}
\end{align}

where $\mathbf{U} \in \mathbb{R}^{|\mathcal{U}| \times k}$ and $\mathbf{V} \in \mathbb{R}^{|\mathcal{E}| \times k}$ are users and relations latent matrices respectively, $R$ is the votes matrix, $\tilde {R}$ is votes approximation matrix and $\|.\|_{\rm {C}}$ is our error measure which leverages users stakes, and is defined as follows:

\begin{equation*}
\|\mathbf{A}\|_{\rm {C}}={\sum _{\mathbf{a}_{i,j} \in \mathbf{A}} I_{i,j}  (s_i \times \mathbf{a}_{i,j}^{2}})
\end{equation*}
$s_i$ is the stake of users $I$, and $I_{i,j}$ is the indicator function, which is 1 if the corresponding element in $R$ is not $*$, and 0 otherwise. In this approach, the majority's power to impose a single honest behavior is limited. That is because the model detects characteristics of each user and relation in the form of latent matrices. As a result, based on the detected latent characteristics of a user, she is even allowed to exhibit an eccentric behavior as long her behavior is consistent, meaning the model is able to find latent characteristics matching her behavior. To put it in other words, a user is considered to have consistent behavior if the model's approximation error is negligible for her. 

One of the noteworthy factors that distinguish our work from common matrix factorization problems is that we do not intend to predict. Instead, we train the model on all of the data, and evaluation of user's behaviors is done by measuring how well their votes fit into the low-rank approximation matrix. If a user's votes deviate significantly from the expected values, the user's reward will dwindle. As we have to use the same data for training and testing, the model can be easily overfitted to achieve smaller error. Nevertheless, we know that overfitted models are not accurate representations of real-world behaviors. In order to prevent overfitting, we need to regularize the model. We follow the regularization approach proposed in \cite{FunkSVD}, and as a result, the final optimization problem is:

\begin{equation} \label{eq:RSBE}
\|\mathbf{R}-{\tilde {\mathbf{R}}}\|_{\rm {C}} + \lambda \times(\|\mathbf{U}\|^2_{fro} + \|\mathbf{V}\|^2_{fro})
\end{equation}

Where $\lambda$ is a regularization parameter set by the network. We call the formula RSBE (Regularized Stake Based Error), denote the first part of it as approximation error, and the second part of it as regularization term.

To Further encourage honest behavior, users are rewarded network tokens based on their behavior. To quantify the behavior of user $i$, we define consistency score:

\begin{equation}
c(i) = \frac{{\sum_{j=1}^{|\mathcal{E}|} I_{i,j}}}{ {\sum_{j=1}^{|\mathcal{E}|} I_{i,j}| \mathbf{r}_{i,j} - \tilde{\mathbf{r}}_{i,j}|^{2}}}
\end{equation}

In this version, the trusted manager of the network fulfills the responsibility of calculating users' consistency scores and will pay rewards to users. Each user's reward is directly related to his consistency score. In the next version, we remove the trusted manager and explore the network operation in a blockchain environment.

\subsection{Version 2} \label{sec:version-2}

In this version, the network operation is explained in a trustless, decentralized environment, meaning blockchain. After the removal of the centralized manager, the first question that comes to mind is that who will carry out the manager's responsibilities. The following subsection answers this question.
\subsubsection{Roles}

The removal of the centralized trusted manager introduces several new roles to the network. Apart from users who stake their tokens and vote on relations, we will need a role that performs the matrix factorization, and a role that checks the validity of the operation. Besides, there is a need for a role that guarantees the data required for validation is available. We define four roles in the network. A brief description of each role is provided below:

\begin{itemize}
	
	\item Staker: in order to play an effective role in the matrix factorization algorithm and gain rewards, stakers lock some of their tokens in the network, and vote on the relations between news items.

	\item Solver: solvers are responsible for solving the matrix factorization problem and committing their results to the main-chain afterward. The solver with the least error is appointed as the winner and is subjected to fraud proofs.
	
	\item Validators: validators can check the integrity of the matrix factorization results and generate fraud proofs in case of a fraudulent solution.
	
	\item Attester: to make sure that enough data is available for validators for making fraud proofs, randomly selected attesters are required to attest to the availability of the data. If most of the attesters attest to the availability of the data, the data is considered available.
	
To contribute to the network as attester or solver, users are required to lock a specific amount of their tokens as a bond. If these users act fraudulently and it is proved on the main-chain, their bond is forfeited and is rewarded to the prover. The details of each role's performance will be delineated in the following sections.

\end{itemize}

\subsubsection{Factorization Algorithm}

The users who participate in the network as solvers are responsible for performing the matrix factorization and committing their results. One of the main requirements of the solvers' jobs is that it should be verifiable. Meaning, validators should be able to validate whether the solver has done his job correctly, and create a fraud proof in case of fraudulent behavior. With this in mind, there are two approaches that can be adopted toward selecting the factorization algorithm for our network: 

\begin{enumerate}
	\item Using a fixed and specific algorithm that all solvers should use.
	\item Have solvers using an arbitrary algorithm and select the solution with the least error.
\end{enumerate}

In the first approach, validation is done by running the algorithm and checking whether the results match the solver. The advantage of this approach is that because of its fixed nature, it easily fits into the truebit verifiable computation model \cite{teutsch2018truebit}, which provides us with a scheme for how to create fraud proofs. On the downside, there are several algorithms that can be utilized to solve the matrix factorization problem \cite{zhou2008large,mnih2008probabilistic,FunkSVD} , and it is likely that newer and more efficient algorithms are yet to be proposed in the future. By fixing the algorithm, this approach deprives solvers of utilizing the state of the art algorithms. As a result, there will be no competition between the users to reach the best result, and all of them will run the same code. Furthermore, updating the algorithm requires an update on the entire network.

On the other hand, using the second approach, solvers can use their desired algorithms to reach the least error, which enables the network to benefit from the latest and best factorization algorithms without imposing an update on the entire network. Nevertheless, using this approach comes at the price of having a more complex validation protocol since we need a more general validation mechanism. This is the approach that we will take in this paper. 

In the next section, we propose a scheme for validation of the results that are provided by solvers following this approach.

\subsubsection{Layer-2 Computation Validator}

Limited computation power is one of the fundamental issues of blockchain that should be accounted for while designing a blockchain-based application. There have been several proposals that increase the processing power of blockhain by changing the core protocols \cite{luu2016secure,eyal2016bitcoin,bagaria2018deconstructing}  and therefore create a new type of blockchain. These proposals are complex and contain several open problems. In order to avoid protocol change and take advantage of the secure and straightforward consensus mechanism of classic blockchain platforms, which is enforced by the very high number of users of these platforms, layer-2 scalability solutions are proposed. These solutions enable us to achieve scalability by using an existing blockchain as a decentralized truthful entity that can judge between parties in case of conflicts. Inspired by truebit \cite{teutsch2018truebit}, we use a layer-2 interactive protocol for on-chain validation of the computations that are done off-chain. The protocol is further discussed in the following section.

\subsubsection{Validation and Fraud Proofs}
\label{Validation}

To have a layer-2 scalability solution, a smart contract should be deployed on the blockchain. This smart contract contains the rules of the network, and given the required data, can judge the disputes. We call this smart contract the \textit{judge smart contract}. After selecting the solver with the least RSBE as the winner, he is required to commit the results on the blockchain through the \textit{judge smart contract}. For a specific period of time, validators can check the submitted results and generate fraud proofs for invalid submissions. After receiving the required data, the smart contract judges between the solver and validator, which will lead to a punishment for the fraudulent actor. Checking the results by validators is done by multiplying the latent matrices committed by the solver, calculating its error using formula \eqref{eq:RSBE}, and comparing it with the committed error.
Before we jump into the fraud proof protocol, we need to introduce a new kind of Merkle tree which provides us with a method for summarizing result .

\subsubsection{Square Sum Annotated Merkle Tree}

In a blockchain environment, the state is replicated on all of the nodes of the network. This high amount of replication increases the cost of saving data on the blockchain. Hence, utilizing techniques to reduce the space requirements of a blockchain-based application is one of the integral parts of the application design. One of the most popular ways of data summarization is using Merkle trees. A Merkle tree \cite{merkle1987digital} enables us to produce a verifiable summary of leaf values in a tree by submitting a single value, which is the root of the tree. In our case, instead of trees, we need to commit matrices. Furthermore, no only the matrix elements, but also the square sum of elements should be present in the resulting verifiable summary. To answer these requirements, we propose a new kind of Merkle tree for matrices called Square Sum Annotated Merkle Tree (SSAMT), which is based on Merkle aggregation \cite{crosby2009efficient}.

Each node's label in a SSAMT is a tuple $(s,h)$. For leaves, $s$ is equal to the value of a matrix element, and $h$ is the hash of it. Let $I$ be an interior node, $H$ be a cryptographic hash function, $I.left$ be the left child of $I$, and $I.right$ be the right child of $I$. We compute the label of node $I$ using formulas: $I.s = I.left.s + I.right.s$, and $I.h = H( I.s || I.left.h || I.right.h)$, where $||$ is concatenation operator.

\begin{figure}[t]	
	\includegraphics[width = .9\textwidth]{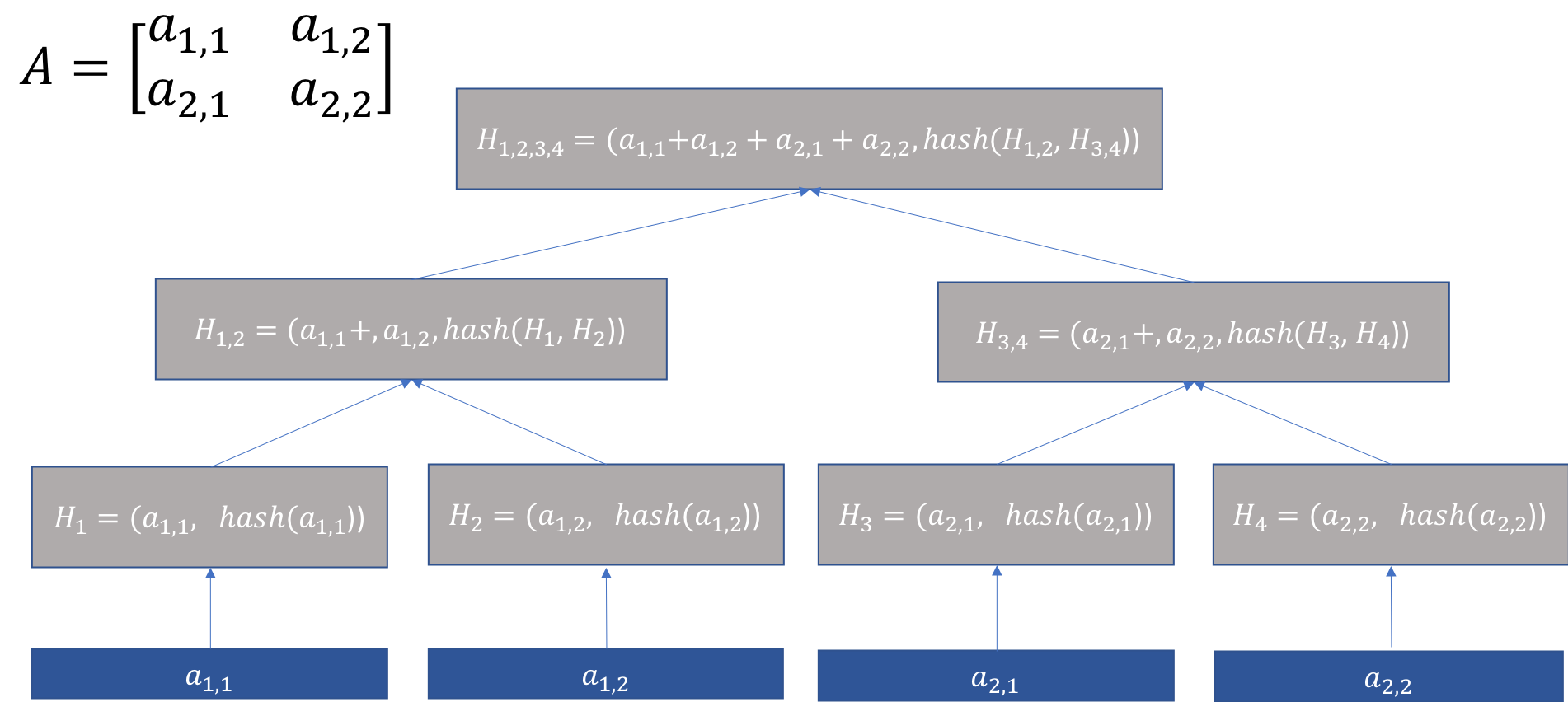}
	\centering
	\caption{
		SSAMT example.
	}
	\label{fig:SSAMT}
\end{figure}

The tree construction procedure starts from the rows of the matrix, and each row turns into a subtree where the elements of the row correspond to the leaves of the subtree. Then, these subtrees are connected following the procedure denoted for the construction of the interior nodes and yield a single tree. This annotated tree enables us to make proofs that enforce the sum of the matrix elements. Figure \ref{fig:SSAMT} shows an example of SSAMT.

Proofs of SSAMT are similar to Merkle proofs \cite{merkle1987digital}. For a matrix element $\mathbf{x}_{i,j}$ the Merkle proof contains the element and siblings of all of the nodes in the path from $\mathbf{x}_{i,j}$ to the root. For a matrix row $\mathbf{X}_{i}^T$, the Merkle proof contains elements of $\mathbf{X}_{i}^T$ and siblings of all of the nodes in the path from the root of the subtree--which comprises the row elements--to the root of the tree. By building an SSAMT, we create a verifiable proof of owning a matrix with a specific sum of the squares of elements.

\begin{figure*}[t]
	\centering
	\subfloat[proof for element $\mathbf{a}_{1,1}$]{\includegraphics[width=.45\linewidth]{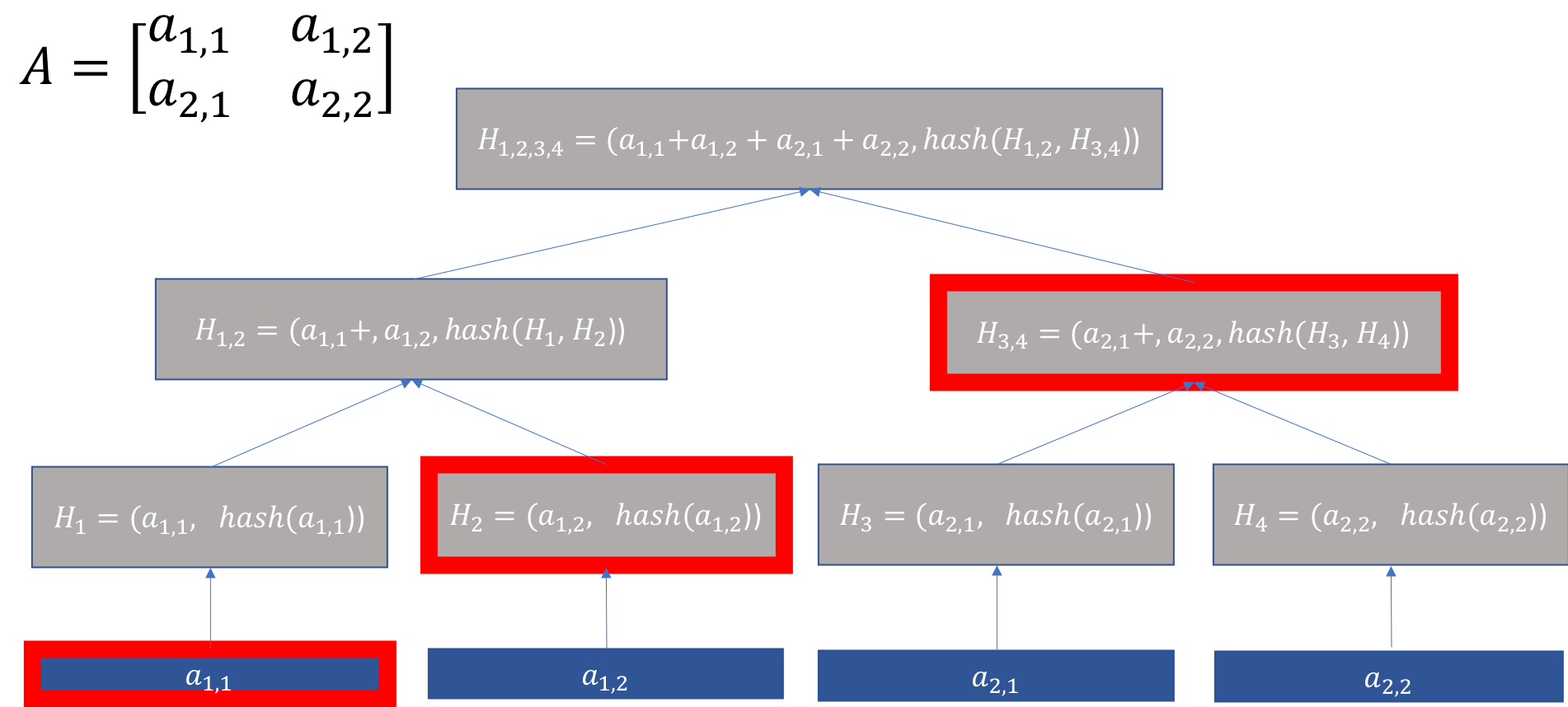}  \label{fig:SSAMT-proof-element}}
	\subfloat[proof for  row $\mathbf{A}^T_{1}$]{\includegraphics[width=.45\linewidth]{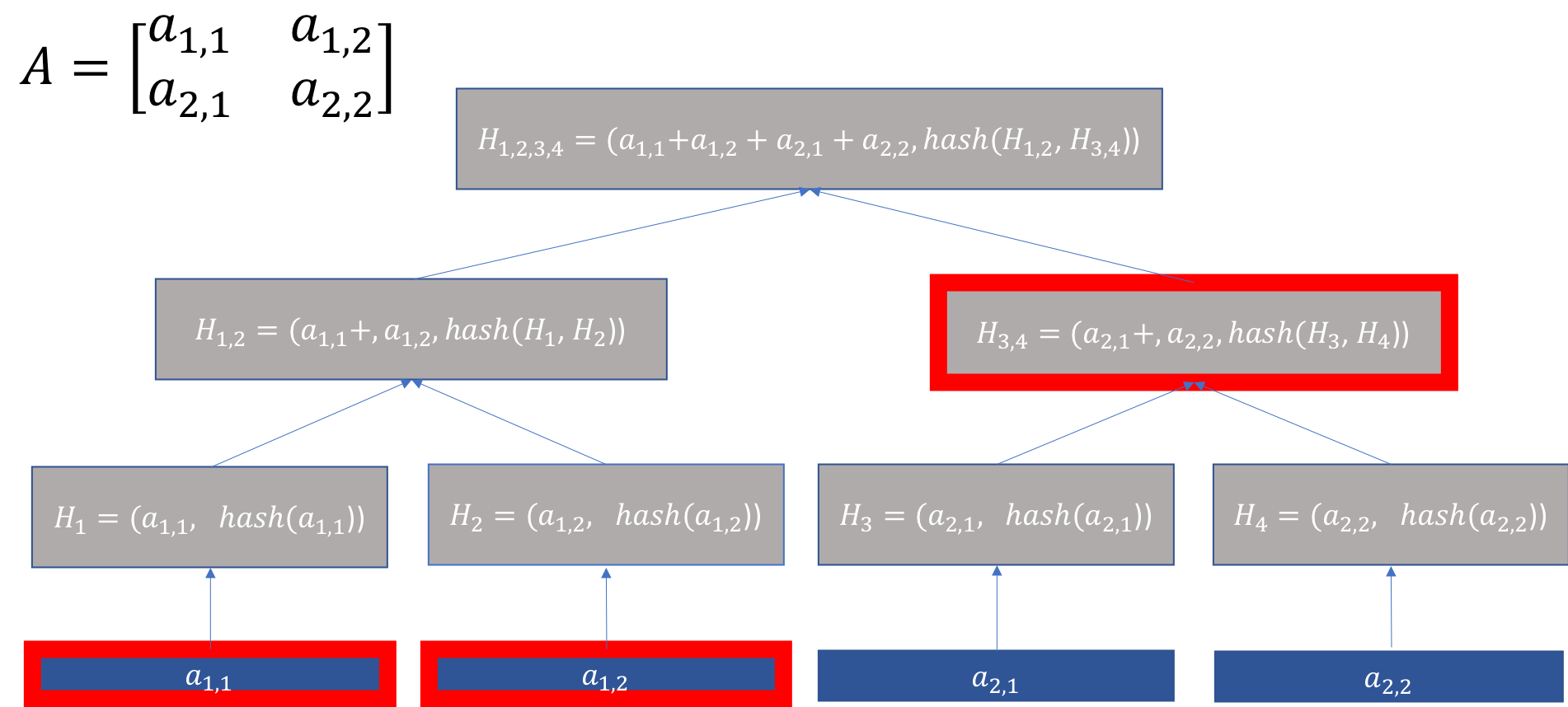}  \label{fig:SSAMT-proof-ًrow}}
	\caption{
		SSAMT proofs
	}
	\label{fig:SSAMT-Proof}
\end{figure*}

\subsubsection{Branch Traverse Procedure}

We only commit the root of an SSAMT or a Merkle tree on the blockchain. If there is a fault in the underlying tree construction or its leaves, the \textit{judge smart contract} will not be able to detect the fault and judge based on the available info. Therefore, we need a procedure to reach the leaves of the tree from its root, and provide the \textit{judge smart contract} with the required information for making a judgment. We call this procedure, which is similar to the approach mentioned in \cite{POC,POC-game}, branch traverse procedure. In this procedure, we have two parties: (1) prover, who has committed a tree root, and (2) verifier, who intends to prove that the tree construction is faulty or a leaf's value is invalid. The procedure consists of multiple steps as follows:

\begin{enumerate}[label=Step \arabic*]
	\item The prover provides the children of the root.
	\item The verifier selects the node which is the root of the subtree having an invalid node. 
	\item The prover submits the children of the selected node.
	\item Steps 2 and 3 are repeated until reaching a leaf.
\end{enumerate}

If the tree construction is faulty, by traversing an invalid branch, an invalid node and its children will be submitted to the blockchain. Having these data, the \textit{judge smart contract} can detect the faulty construction, and punish the prover. If the verifier intends to prove that a leaf is invalid, after the end of the branch traverse procedure, the \textit{judge smart contract} will have the leaf's value, and the verifier should submit his proof for its invalidity.

\subsubsection{Fraud Proof Protocol} \label{ssec:Fraud_Proof}

In this section, we explain how a validator can prove to the \textit{judge smart contract} that a result, which is submitted by a solver, is faulty. As mentioned before solvers should find matrices $\mathbf{U}$, and $\mathbf{V}$ such that RSBE, formula \eqref{eq:RSBE}, is minimized. We define error matrix $\mathbf{EM} = [\mathbf{em}_{i,j} ] \in \mathbb{R}^{|\mathcal{U}| \times |\mathcal{E}|}$ as:

\begin{equation*}
\mathbf{em}_{i,j} =  \sqrt{s_i\times (\mathbf{U}_i^T . \mathbf{V}_j^T - \mathbf{r}_{i,j})^{2} }
\end{equation*}

where $\mathbf{U}_i^T$ is the $i$th row of users' latent matrix, $\mathbf{V}_j^T$ is the $j$th row of relations' latent matrix and $s_i$ is the stakes of user $i$. In order to make the submitted result's verification feasible on blockchain, the protocol asks the solver to make a commit containing several pieces of data including (1) The SSAMT root of users' latent matrix $SR_U$, (2) The SSAMT root of relations' latent matrix $SR_V$, and (3) the SSAMT root of the error matrix $SR_{EM}$.


After the commitment, validators can check the integrity of the solution and submit a challenge if the results are invalid. To do so, they should first multiply the latent matrices to compute their own error matrix $\mathbf{VEM}$ and its SSAMT root $SR_{VEM}$. Then they compare the results with the commitments, and if there is a contradiction, they send a challenge transaction to the \textit{judge smart contract}. The purpose of the challenge transaction is to prove to the \textit{judge smart contract} that the RSBE claimed by the solver is not reachable using the committed matrices.

The source of the contradiction between the solver and a validator's computations can be in the regularization term or approximation error (error matrix). Contradiction on the regularization term means $\mathbf{U}$ or $\mathbf{V}$ do not result in the claimed Frobenius norm. This case is covered by the data availability protocol (next section). Contradiction on the error matrix means that the error matrix is not valid given the submitted $\mathbf{U}$ and $\mathbf{V}$ ($\mathbf{EM} \neq \mathbf{VEM}$). Suppose element $\mathbf{em}_{i,j}$ is invalid. In this case, the validator sends a challenge transaction that starts a branch traverse procedure on the tree with root $SR_{EM}$. When the branch traverse procedure is done to the leaf corresponding to $\mathbf{em}_{i,j}$, the solver sends $\mathbf{U}_i^T$, $\mathbf{V}_j^T$, and their SSAMT proofs (which correspond to $SR_U$ and $SR_V$ committed by the solver) to the \textit{judge smart contract}.

Having these data, the \textit{judge smart contract} can calculate the resulting error and decide whether the solver committed an invalid result or the validator's fraud claim is invalid. In the end, the faulty actor is punished, and the honest actor is rewarded.

\subsubsection{Data availability} \label{sssec:Data availability}

As mentioned previously, to validate a solver's submission, validators should multiply latent matrices $\mathbf{U}$ and $\mathbf{V}$, and check whether the claimed result is achievable. These matrices have a row of size $k$ for each user and each relation, and consequently, grow linearly as the network grows, making them too large to be stored on the blockchain. As a result, we only store their SSAMT roots on the blockchain, and if there is a fault in the SSAMT construction or elements of the matrix, it can be proved to the \textit{judge smart contract} through the fraud proof protocol. The prerequisite of this protocol is that the validators have access to the latent matrices and are able to check the validity of solvers' results. The fraud proof protocol by itself is not enough to guarantee the access of validator to the latent matrices. We need a method to provide a guarantee that the data is available to validators for download. In order to achieve this goal, we incorporate the proof-of-custody scheme proposed by Buterin in \cite{POC,POC-game} into our fraud proof protocol.

In this scheme, an attester role is introduced, and users affiliated with this role are required to attest to the availability of the data needed by the validators. Since downloading and checking data validity is more expensive than simply attesting to every data without downloading, the challenge here is how to validate the attestations and prevent the attestations to unavailable data. To do so, after the selection of the solver, attesters are required to compute a secret salt, $s$ by signing the a public parameter $p$ using their private key (we will assign the value of $p$ in the next version). After the commitment by the solver, the attesters should attest to the availability of the matrices corresponding to the SSAMT roots $SR_U$, and  $SR_V$. Let $||$ be the concatenation operator, and denote the matrix to be attested at as $D$. The attestation is constructed by splitting the data into $n$ chunks, $D[1],D[2],...,D[n]$, for each chunk computing $C[i]=D[i] || s $, and making a Merkle tree of $C[i]$ chunks. We denote the attester's Merkle tree as $MT_A$. In the case of our verification protocol, $D$ is a matrix, and each chunk of it is a matrix element. The attester then commits the resulting Merkle root $MR_A$ to the main-chain. When the attestations are complete, attesters are required to reveal their $s$. Validators can check the validity of the attestation, and send a challenge in case of malicious behavior. Using this scheme, two cases are possible:

First, the data is available. In this case, if an attester attests to the data without following the protocol when the secret salt is revealed, the misbehavior can be detected by validators. To validate an attestation, validators should use attester's secret salt, create their own Merkle tree $MT_V$, and compare the resulting root $MR_V$ with $MR_A$ committed by the attester. If they are not the same, the validator can issue a challenge, which leads to the start of a branch traverse protocol, where the attester plays the role of the prover, and the validator plays the role of the verifier. The branch traverse procedure ends to a leaf, which the validator believes has an invalid value. Next, the attester should submit two pieces of data: (1) $D[i]$ corresponding to the leaf, and (2) an SSAMT proof showing $D[i]$ is included in the matrix corresponding to SSAMT root committed by the solver, at the beginning of the fraud-proof protocol. If the procedure continues until the end, it means that the attestation was correct, and the validator gets punished. On the other hand, if the attester is unable to provide a valid response in each step, he is deemed fraudulent and punished.
Based on the provided information, the \textit{judge smart contract} can rule whether the validator's challenge was valid or not, and as a result, rewards are paid, and punishments are carried out.

Second, the data is unavailable. In this situation, the validators will be caught in a dilemma \cite{data-availability}, meaning based on the fact that they do not have access to the data, they cannot infer that neither do the attesters. In this scenario, if a validator challenges the attester and the attester having had the data had made a valid attestation, the validator will lose the challenge and get punished. In order to overcome this problem as suggested in \cite{POC-game}, we allocate a slot of free challenges in each epoch, which can be filled by validators, and allows them to ask an attester to submit $C[x]$ where $x$ is an arbitrary index. In response to this challenge, the attester is required to send $C[x]$ together with its Merkle proof (corresponding to the Merkle root committed by the attester), and $D[x]$ together with its SSAMT proof (corresponding to a SSAMT root committed by the solver).

In the next version, the described actions and protocols of the network will be put into a schedule, and the timing will be declared.

\subsection{Version 3} \label{sec:version-3}

In this version we conclude the details of the proposed network by delineating the order of the actions in the network and their timing constraints. 

We divide the network's operation is into intervals named cycles. During each cycle, users stake their tokens and votes on the news items relations. Users are required to submit at least ten votes during the cycle to be taken into account in the incentive mechanism. At the end of the cycle, users are rewarded based on the analysis of their behavior during the cycle, which is done by the incentive mechanism.

Each cycle includes two overlapping phases: the voting phase and the evaluation phase. In the voting phase, users willing to contribute in the network as stakers, commit the hash of their desired value concatenated with an arbitrary secret and vote on the relations. Committing stakes can be done at any time during the voting phase. 

The evaluation phase consists of multiple stages, as follows:

\begin{enumerate}
	\item Stakers reveal the secret and value used in their hashed commitment, and lock tokens equal to the revealed value as their stakes. Users who fail to reveal their commitment will be ignored.
	\item attesters lock a specific amount of tokens as a bond.
	\item Volunteer solvers start computing latent matrices and commit SSAMT roots of their results.
	\item The solver proposing the best result is appointed as the winner.
	\item Randomly selected attesters, attest to data availability.
	\item Attesters reveal their secret salt.
	\item Challenges are submitted.
	\item Challenges are judged.
	\item Stages 2-8 are repeated until there is no challenge submitted in step 7.
	\item rewards are paid.
\end{enumerate}

As mentioned in Section \ref{sssec:Data availability}, attesters should sign a public parameter $p$ using their private keys. Let $cn$ be a 256-bit number representing the number of the attestation's cycle, and $rn$ be a 256-bit number representing the number of times that the evaluation phase has been repeated in that cycle. In this situation, attesters should use $p=hash(cn || rn )$.

To receive her reward, the user $i$ should submit the model's computed error for her to the smart judge contract. This is done by sending the sum of the elements of row $\mathbf{E}_i^T$. The user can withdraw her reward instantly by submitting an SSAMT for $\mathbf{E}_i^T$. The other option is to wait two weeks, and if no validator submits a fraud proof for this user's claimed error, the user can withdraw her reward.

\begin{figure}
	\centering
	\includegraphics[width=0.8\linewidth]{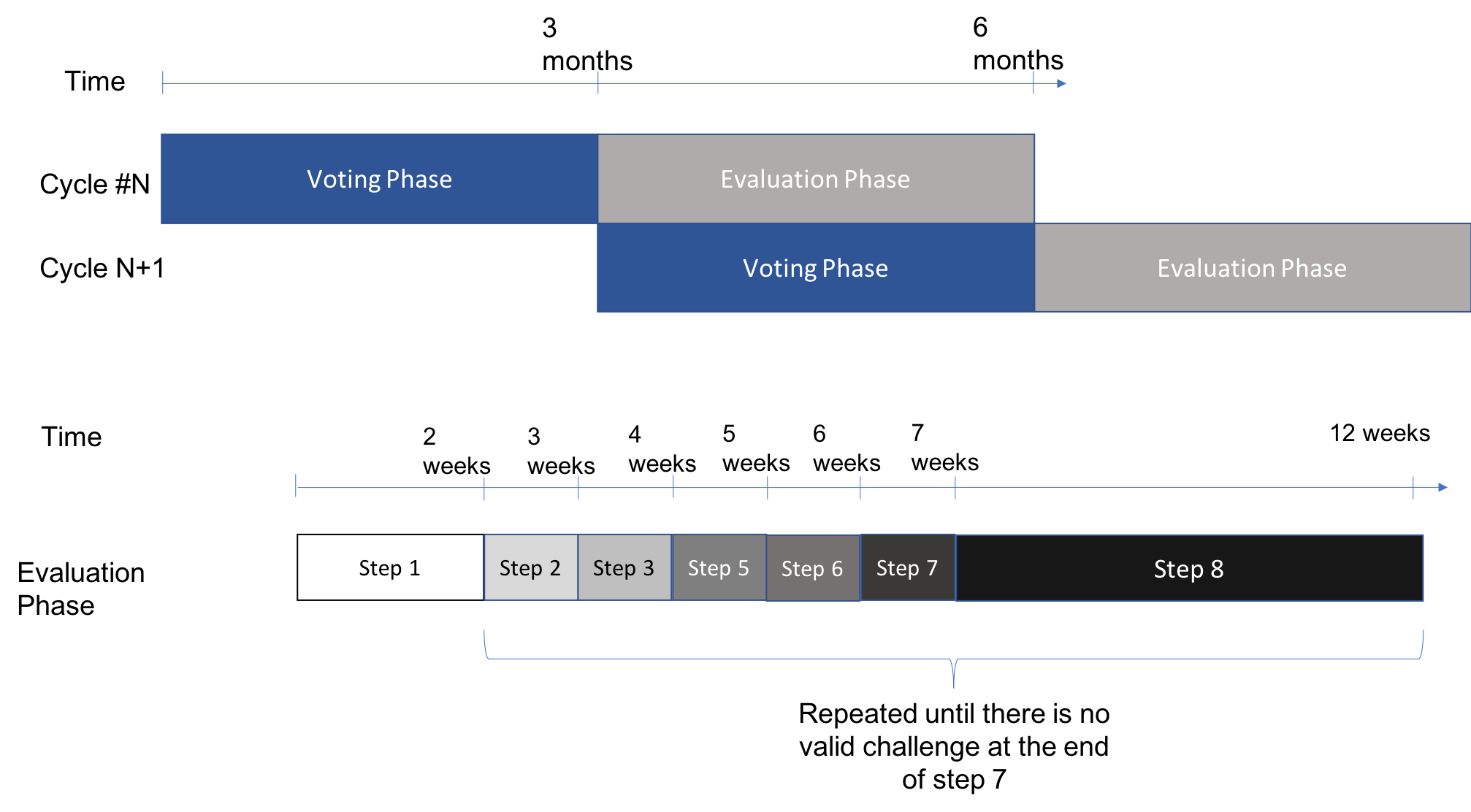}
	\caption{A Cycle in the network - step 4 because it is executed instantly after step 3, and step 8 because it does not stall the network from proceeding to the next cycle, are not depicted in the figure}
	\label{fig:fig3}
\end{figure}

The commit-reveal process of staking makes the model unpredictable and prevents users from computing the most profitable way to vote before the end of the voting phase. Figure \ref{fig:fig3} shows phases and their timings in the case that no challenge is submitted during the evaluation phase. 

A cycle's voting phase can be concurrent with multiple evaluation phases from previous cycles because although the voting phase is repeated every three months, the duration of the evaluation phase can vary based on the number of times the results are challenged.

\section{Incentive Mechanism Evaluation} \label{sec:evaluation}

In this section, we evaluate the performance of the proposed matrix factorization solution. Because of the novelty of our approach regarding voting on relations instead of the items, and the existence of adversary users, there is no real-world dataset available that could serve our purpose. Therefore, synthesizing required data was the best option we were left with. In order to evaluate the robustness of our algorithm with respect to adversarial behaviors, we applied the proposed approach to a comprehensive range of synthetic datasets representing different environments that the network may operate in. In this section, first, the analysis of fraud proof and data availability protocols are given. Then, we suggest an algorithm for implementing the proposed matrix factorization solution. Next, we describe data datasets, and at last, the evaluation of the suggested algorithm is presented.

\subsection{Analysis of Fraud Proof and Data Availability protocols}

Since in a blockchain network, the blocks are replicated on all of the nodes, space is expensive and should be taken into account while analyzing algorithms. In this section, we will analyze the time and space required by the proposed fraud proof protocol. The analysis ignores auxiliary data sent in the transactions, including addresses and signatures. Let us name each repetition of steps 2 and 3 of the branch traverse procedure a around. The number of rounds is equal to the depth of the tree be traversed. Therefore, if a tree has $x$ leaves, the number of rounds for the branch traverse procedure will be $\log_{}(x)$. As a result, the total number of steps is:

\begin{equation*}
2 + 2 \log_{}(x) = \mathcal{O}(\log_{}(x))
\end{equation*}

In each round, two nodes are submitted by the prover. If we ignore 1 bit sent by the verifier for selecting a child, the sum of the data submitted during the branch traverse procedure is:

\begin{equation*}
2\log{}(x) \times  ns \quad \text{bytes}
\end{equation*}

Where $ns$ denotes the size of the tree nodes. Each SSAMT node is a tuple $(s,h)$ where $s$ is a 32-byte integer, and $h$ is a 32-byte hash. So, for a SSAMT we have: $ns=64 \; \text{bytes}$. In a Merkle tree, on the other hand, each node is a single 32-byte hash, and we have $ns=32 \; \text{bytes}$. The Fraud proof schema requires a commitment from the solver, which contains three SSAMT roots for a total size of $192$ bytes. If a verifier submits a challenge for the results committed by the solver, a branch traverse procedure starts on an SSAMT with depth $\log_{} (|\mathcal{U}| \times |\mathcal{E}|)$. At the end of the procedure, the solver should submit (1) a row of the users' latent matrix together with its SSAMT proof with a total size of $k \times 32 + log(|U|) \times 64$ bytes, and (2) a row of the relations' latent matrix together with its SSAMT proof with a total size of $ k \times 32 + log(|V|) \times 64$ bytes. Therefore the total size of the data submitted to settle the challenge in the fraud proof protocol is:

\begin{equation*}
128\log_{}(|\mathcal{U}| \times |\mathcal{E}|) + \\
( \log_{}(|\mathbf{U}|) + \log_{}(|\mathbf{V}|)+ k) \times 64 \quad \text{bytes}
\end{equation*}

Each attester should commit two 32-byte Merkle roots at each cycle. In case a verifier submits a challenge for an attester's commitments, at the end of the branch traverse procedure, the solver should submit a user's latent matrix row, and its SSAMT proof. Therefore the total size of the data submitted to settle the challenge in the data availability protocol is:

\begin{equation*}
128 MV + 32 \quad \text{bytes}
\end{equation*}

Where $ MV = |\mathbf{U}| \text{or} |\mathbf{V}|$ based on the Merkle root which is challenged.

The number of rounds in the branch traverse procedure can be reduced by requiring the verifier to send more depth of its tree in each round, nevertheless, with the cost of more data sent in each round. For example, if in step 3, the verifier sends each node's grandchildren too, the number of the required rounds will be halved, but in each round, six nodes are sent instead of two.

\subsection{Matrix Factorization Algorithm} \label{ssec:MF-algorithm}

Formula \eqref{eq:RSBE} shows the objective function that should be minimized. To do so, we use the simple approach proposed in \cite{FunkSVD}, which has great performance on sparse matrices with missing data. In this approach, the overall error is minimized by minimizing the error for each element using gradient descent. Let $\mathbf{e}_{i,j}$ denote the approximation error for vote of user $i$ to relation $j$. Therefore, $\mathbf{e}_{i,j}$ can be formulated as:

\begin{equation*}
\mathbf{e}_{i,j} = s_i \times ( \mathbf{r}_{i,j} - \sum_{f=1}^{k} \mathbf{u}_{i,f} . \mathbf{v}_{j,f})^{2}
\end{equation*}

We define regularized error $\mathbf{re}$ for user $i$ and relation $j$ as:

\begin{equation*}
\mathbf{re}_{i,j} = \mathbf{e}_{i,j} + \lambda (\sum_{f=1}^{k} \mathbf{u}_{i,f}^2 + \sum_{f=1}^{k} \mathbf{v}_{j,f}^2)
\end{equation*}

By taking the derivative of $\mathbf{re}_{i,j}$ with respect to $u_{i,f}$ we have:

\begin{equation*}
\frac{\partial \mathbf{re}_{i,j}}{\partial \mathbf{u}_{i,f}}= -2  \mathbf{e}_{i,j}\mathbf{v}_{j,f} + 2 \lambda u_{i,f}
\end{equation*}

Similarly by taking the derivative of $\mathbf{re}_{i,j}$ with respect to $\mathbf{v}_{j,f}$ we have:

\begin{equation*}
\frac{\partial \mathbf{re}_{i,j}}{\partial \mathbf{v}_{j,f}}= -2  \mathbf{e}_{i,j}\mathbf{u}_{i,f} + 2 \lambda \mathbf{v}_{j,f}
\end{equation*}

By combining the constants into regularization variables, we will have the update rules:

\begin{equation*}
\mathbf{u}_{i,f} = \mathbf{u}_{i,f} + \gamma (\mathbf{e}_{i,j}\mathbf{v}_{j,f} - \beta \mathbf{u}_{i,f})
\end{equation*}
\begin{equation*}
\mathbf{v}_{j,f} = \mathbf{v}_{j,f} + \gamma (\mathbf{e}_{i,j}\mathbf{u}_{i,f} - \beta \mathbf{v}_{j,f})
\end{equation*}

Where $\gamma$ denotes the learning rate, and $\beta$ denotes the regularization variable. To execute this algorithm, matrices $\mathbf{U}$ and $\mathbf{V}$ are initialized randomly, and the algorithm is executed for a specific number of repetitions. We denote this algorithm as SBSVD (Stake Based Singular Value Decomposition).

\subsection{Dataset} \label{ssec:dataset}

Since there was no previous method for generating votes and relations with our desired properties (votes on relations while respecting subjectiveness) we use combination of two methods to synthesize our datasets. One for generating the relations graph, and one for assigning user votings to relations.

\begin{figure}[t]	
	\includegraphics[width = .9\textwidth]{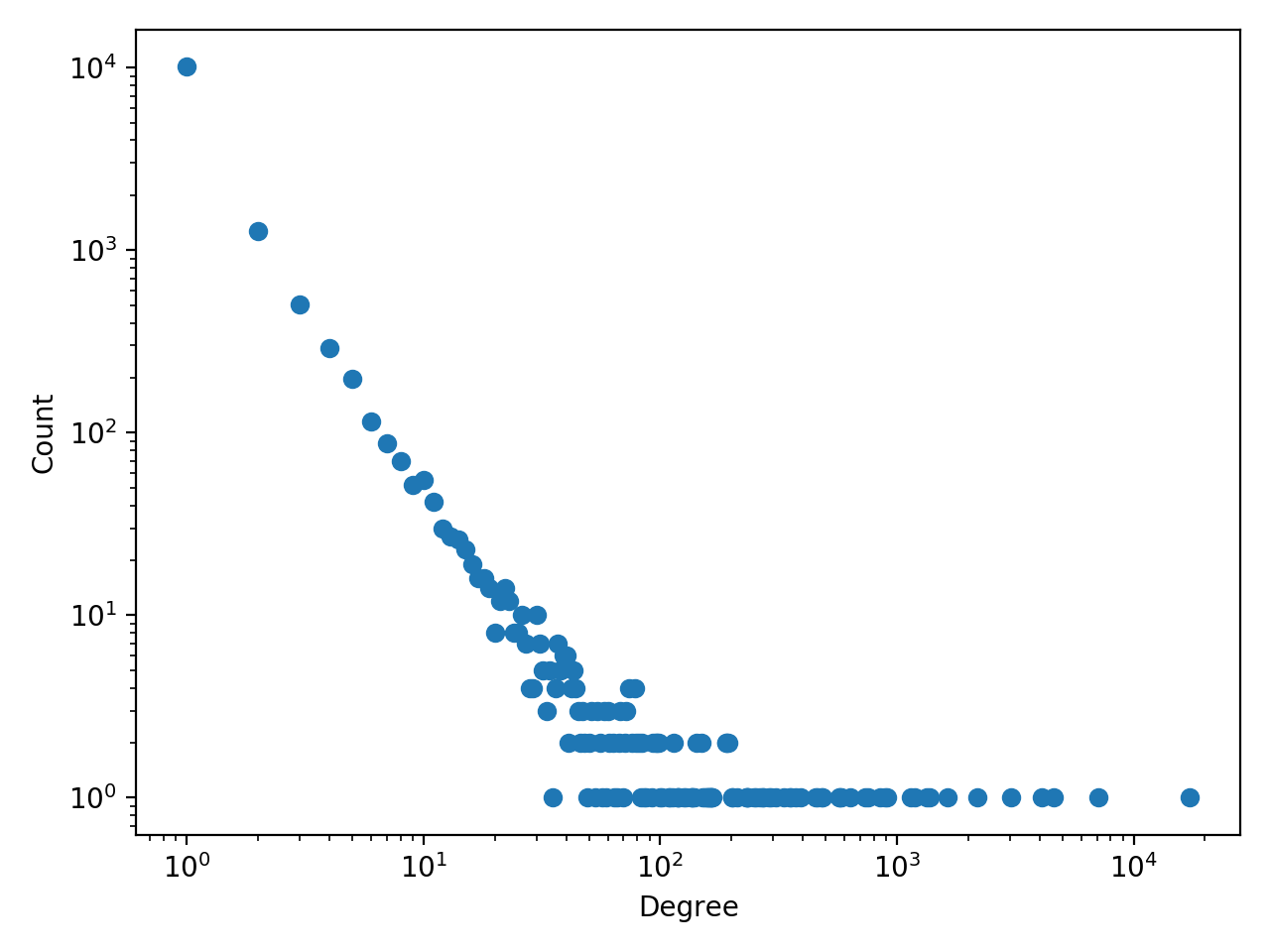}
	\centering
	\caption{
		The degree distribution of the synthesized graph
	}
	\label{fig:degree-distro}
\end{figure}

In order to generate the relations graph we used RTG ( Random Typing Generator ) [...] which creates realistic graphs with respect to several power laws identified in real-word graphs. We utilized RTG with parameters $W=5000$, $k=5$, $\beta=0.6$, and $q=0.4$ and the result was a graph with 17063 nodes, 5000 edges and 1832 components. In order to have a group of closely related news items we picked the biggest component with 13343 nodes and 48102 edges. The resulting graph is multigraph, and there might be multiple edges between two nodes. Each edge represents a relation between two news items.Therefore, based on the formulation presented in section \ref{ssec:graph-fromulation}, $|\mathbf{R }|$ is equal to the number of edge, and $|\mathcal{E}|$ is equal to the number of edges in the corresponding simple graph, which is equal to 22885 for our synthesized graph. We consider the average number of votes for users to be 20. There the number of users is equal to $\frac{|\mathbf{R}|}{20}=2406$. Figure \ref{fig:degree-distro} shows the degree distribution of the final news graph.

In order to synthesize users and votes, we took advantage of the supposition that honest users are expected to have consistent behavior, and as a result, they can be characterized using a few latent features. The same thing applies to relations so, if all of the users are honest, the approximated low-rank matrix, as the result of matrix factorization, will have a negligible error. This is a plausible supposition, backed by the remarkable performance of matrix factorization in a range of different review networks \cite{mnih2008probabilistic, FunkSVD, ortega2016recommending, xu2018novel, jamali2010matrix, zhu2012discovering, ma2008sorec}. Moreover, we add different degrees of noise to our datasets to analyze its performance in the environments that do not satisfy this assumption entirely.

Let $\mathcal(U)(a,b)$ denote the uniform distribution on interval [a,b], and  $\mathcal{N}(\mu, \sigma^2)$ denote the normal distribution with mean $\mu$ and variance $\sigma^2$. We made a low-rank matrix $\mathbf{L} = \mathbf{A}\mathbf{B}^T \in  \mathbb{R}^{|\mathcal{U}| \times |\mathcal{E}|}$ by multiplying two randomly generated matrices $ \mathbf{A} \in \mathbb{R}^{|\mathcal{U}| \times 10}$ and $ \mathbf{B} \in \mathbb{R}^{|\mathcal{E}| \times 10}$ whose elements are drawn from $\mathcal{N}(0,1)$. Then we scaled the matrix elements through multiplying them by a coefficient so that $\max_{\mathbf{l}_{i,j} \in \mathbf{L}} |\mathbf{l}_{i,j}| = 1$. We proceeded to adding noise to the matrix elements using samples drawn from $\mathcal{N} (0, \sigma_v^2)$, where a range of values are assigned to $\sigma_v^2$ to generate datasets with different levels of noise. Next, values are rounded to the closest integer in $\{-1,0,1\}$. In the end, to select a proportion, $ap$, of users as adversaries, each user is considered adversary with the probability $ap$, and it's vote is selected randomly from $\{-1,0,1\}$ with uniform probability.

Now we have a news graph representing relations between news items and a votes matrix, which indicates how each user would vote on a relation. The last step is assigning each relation to a user. Let $\mathcal{P}(\lambda)$ denote the Poisson distribution with rate $\lambda$. Regarding the fact that each user should vote for at least ten times to contribute in the incentive mechanism, and the average number of votes is assumed to be 20, first, for each user we draw samples from $\mathcal{P}(10)$, and add 10 to it, in order to get the number of votes that the user will cast. After determining the number of votes for each user, we randomly assign the graph edges to the users.

\subsection{Parameters}
Several parameters were introduced in the previous section that need to be assigned in order to analyze the proposed algorithm. We divide the parameter into three groups:

\begin{enumerate}
	\item Network parameters: parameter $\lambda$ in formula \eqref{eq:RSBE} is determined by the network. In the following experiment, we consider $\lambda$ to be $0.1$.
	\item Dataset parameters: as explained in the previous section, several distributions are put to use in the process of generating a dataset. The distributions have several parameters. In the previous section, we determined some of the parameters' values, and the rest will be assigned based on the desired characteristics of the final dataset. The dataset parameters that need to be assigned are noise variance  $\sigma_v^2$, and adversary proportion $ap$. In addition, in the following experiment, we consider honest users to stake the maximum number of tokens possible allowed (5 tokens), but the amount of stake for each adversary user, $as$, will be assigned as a dataset parameter.
	\item Algorithm parameters: the algorithm introduced in section \ref{ssec:MF-algorithm} has several parameters, including learning rate $\gamma$, regularization variable $\beta$, number of latent factors $k$, and number of repetitions. Different values for these parameters can result in different errors, and solvers should find the best values which yield the least RSBE. In the following experiment, we consider $\gamma$ to be $0.005$, and the number repetitions to be $1000$. The parameters $\beta$ and $k$ are assigned by the solver, and we denote them as the algorithm parameters.
\end{enumerate}

\subsection{Experiment}
In the section, we analyze the performance of SBSVD algorithm described in section \ref{ssec:MF-algorithm}. First, the application of the algorithm by solvers is explained. Then, we present the analysis of the performance of the algorithm on a range of synthesized datasets, which are generated through the process described in section \ref{ssec:dataset}.

\subsubsection{Application of the Algorithm by Solvers} \label{ssec:algorithm-parameters}

\begin{figure}[t]	
	\includegraphics[width = .9\textwidth]{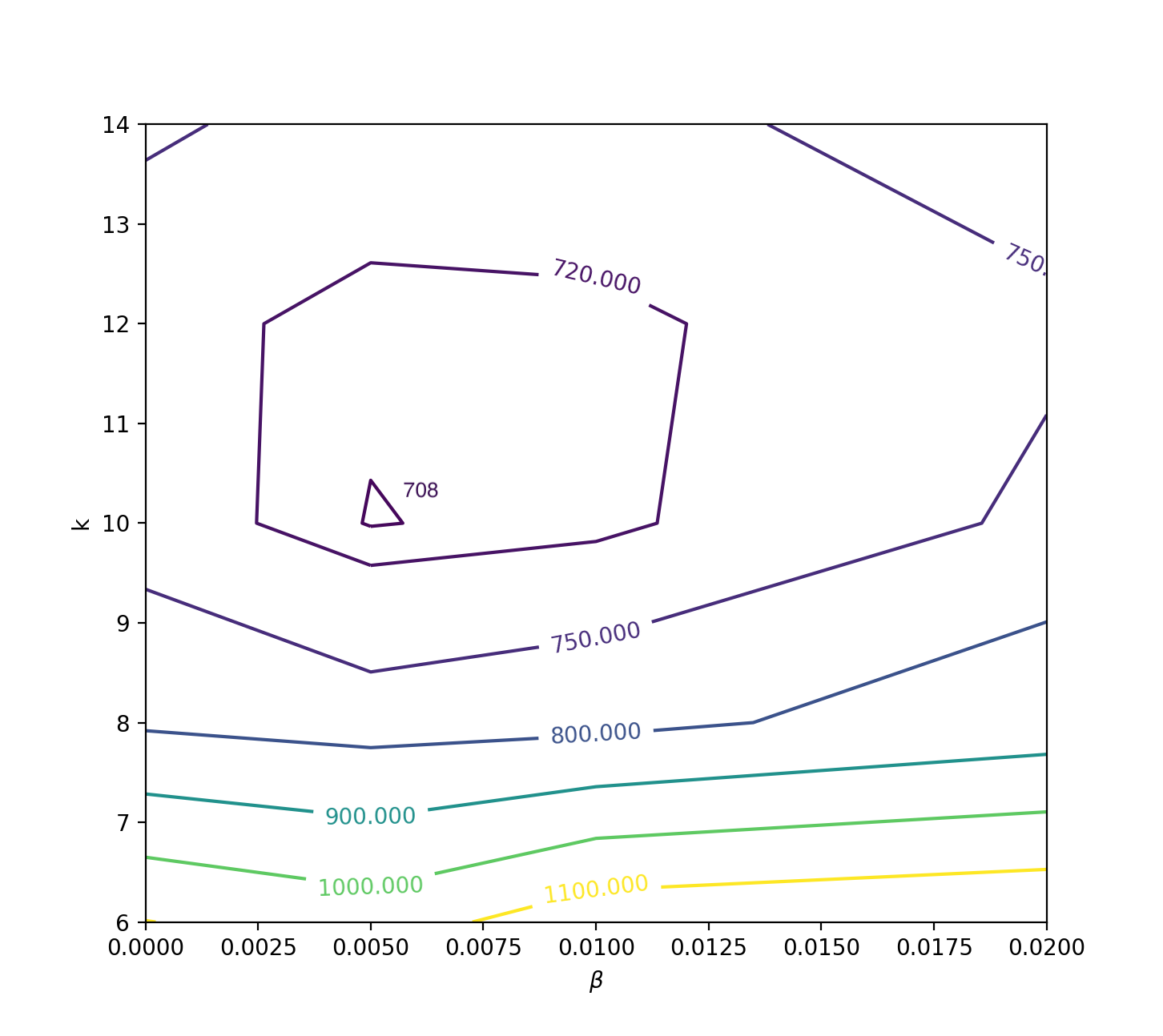}
	\centering
	\caption{
		change in RSBE with respect to parameters $k$, and $\beta$.
	}
	\label{fig:k-and-beta}
\end{figure}

Solvers should try to minimize RSBE. In each cycle, the solver with the least RSBE is appointed as the winner and is rewarded if no valid challenge is submitted by validators. Each solver has the freedom to choose a desired matrix factorization algorithm, one of which is the SBSVD algorithm described in section \ref{ssec:MF-algorithm}. In this section, we outline how a solver can use SBSVD. As an example, consider a dataset generated with parameters $\sigma_v^2=0.1$, $ap=0.1$, and $as=3$. To achieve the least RSBE, the solver should find the best algorithm parameters $k$ and $\beta$. Figure \ref{fig:k-and-beta} shows the resulting RSBE for running algorithm with parameters $k \in \{6,8,10,12,14\}$, and $\beta \in \{0,0.005,0.01,0.02\}$ as a contour plot. Using $k=6$ results in a very high value for RSBE and using higher values for  $k$ can lead to lower values.
Nevertheless, increasing the value of $k$ is not always optimal and will eventually lead to an increase in the value RSBE. The reason is that, because of the overfitting caused by the high values of $k$, the value of the regularization term in formula \ref{eq:RSBE} increases significantly, and leads to the escalation of RSBE. 

Parameter $\beta$ determines the weight of the regularization term, the greater the value of $\beta$, the greater the impact of regularization term on the final error.

In this example the least RSBE is achieved by using parameters $k=10$, and $\beta=0.005$. Therefore, the solver will send the results achieved by these parameters, and if no other solver achieves a lower RSBE he will be appointed as the winner. 

\subsubsection{Evaluation Measure}
In order to evaluate the performance of the SBSVD algorithm, we introduce a new measure called ER (error ratio). Before we get to the definition of ER, we define RMSE for a set of users $\mathcal{S}$ as follows:

\begin{equation*}
RMSE(\mathcal{S}) = \sqrt{\frac {\sum_{i \in |\mathcal{S}|} \sum_{j=1}^{|\mathcal{E}|} I_{i,j}| \mathbf{r}_{i,j} - \tilde{\mathbf{r}}_{i,j}|^{2}}{|\mathcal{S}|}}
\end{equation*}


Let $\mathcal{HU}$ denote the set of honest users, and $\mathcal{AU}$ denote the set of adversary users. We define measure ER as:

\begin{equation*}
ER = \frac{RMSE(\mathcal{AU})}{RMSE(\mathcal{HU})}
\end{equation*}

\subsubsection{Results}
\begin{figure*}[t]
	\centering
	\subfloat[ $ap= 0.1$ ] {\includegraphics[width=.45\linewidth]{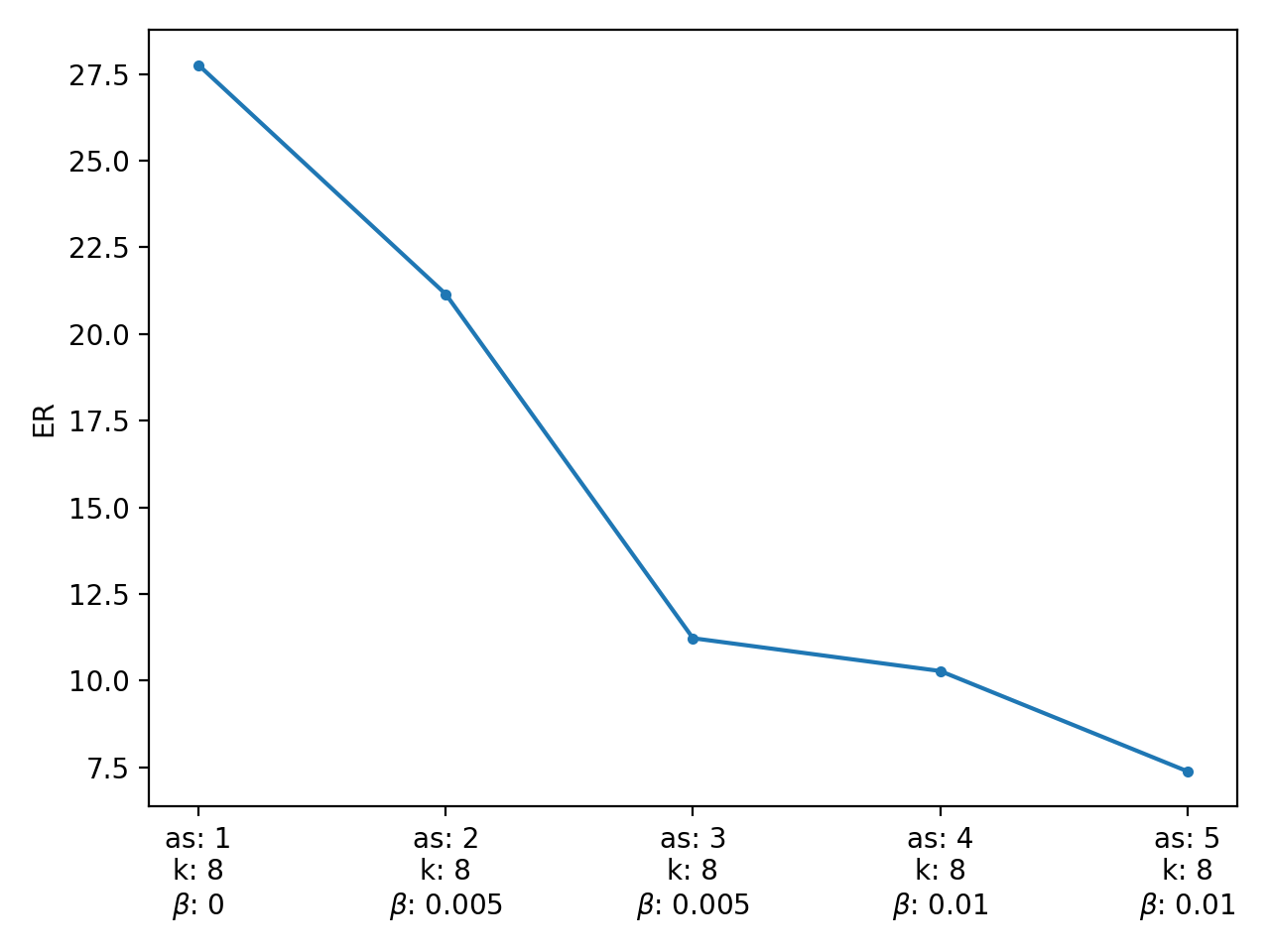}  \label{fig:noise-0-rel-0-1}} 
	\subfloat[ $ap= 0.3$ ] {\includegraphics[width=.45\linewidth]{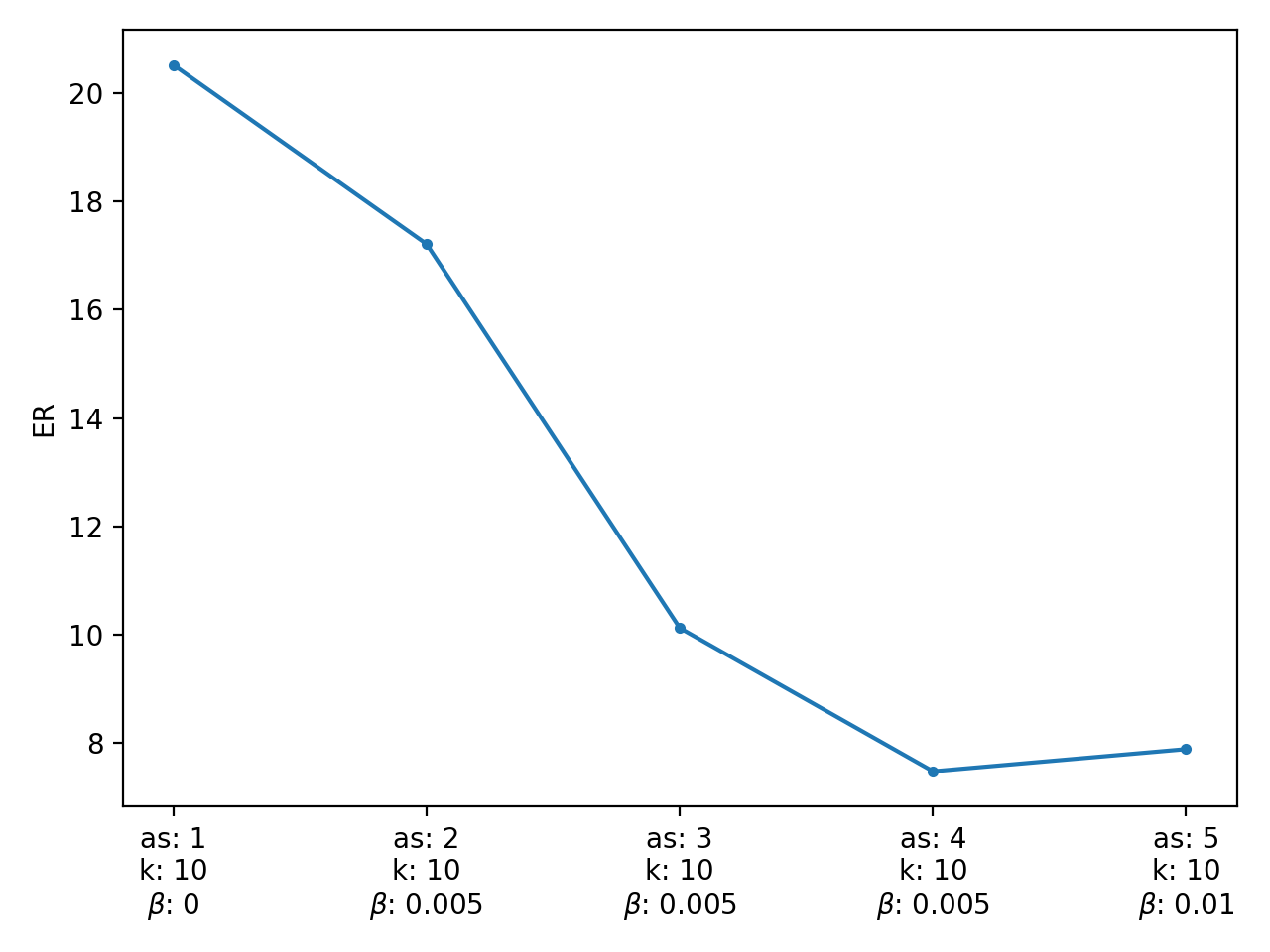}  \label{fig:noise-0-rel-0-3}} \\
	\subfloat[ $ap= 0.5$ ] {\includegraphics[width=.45\linewidth]{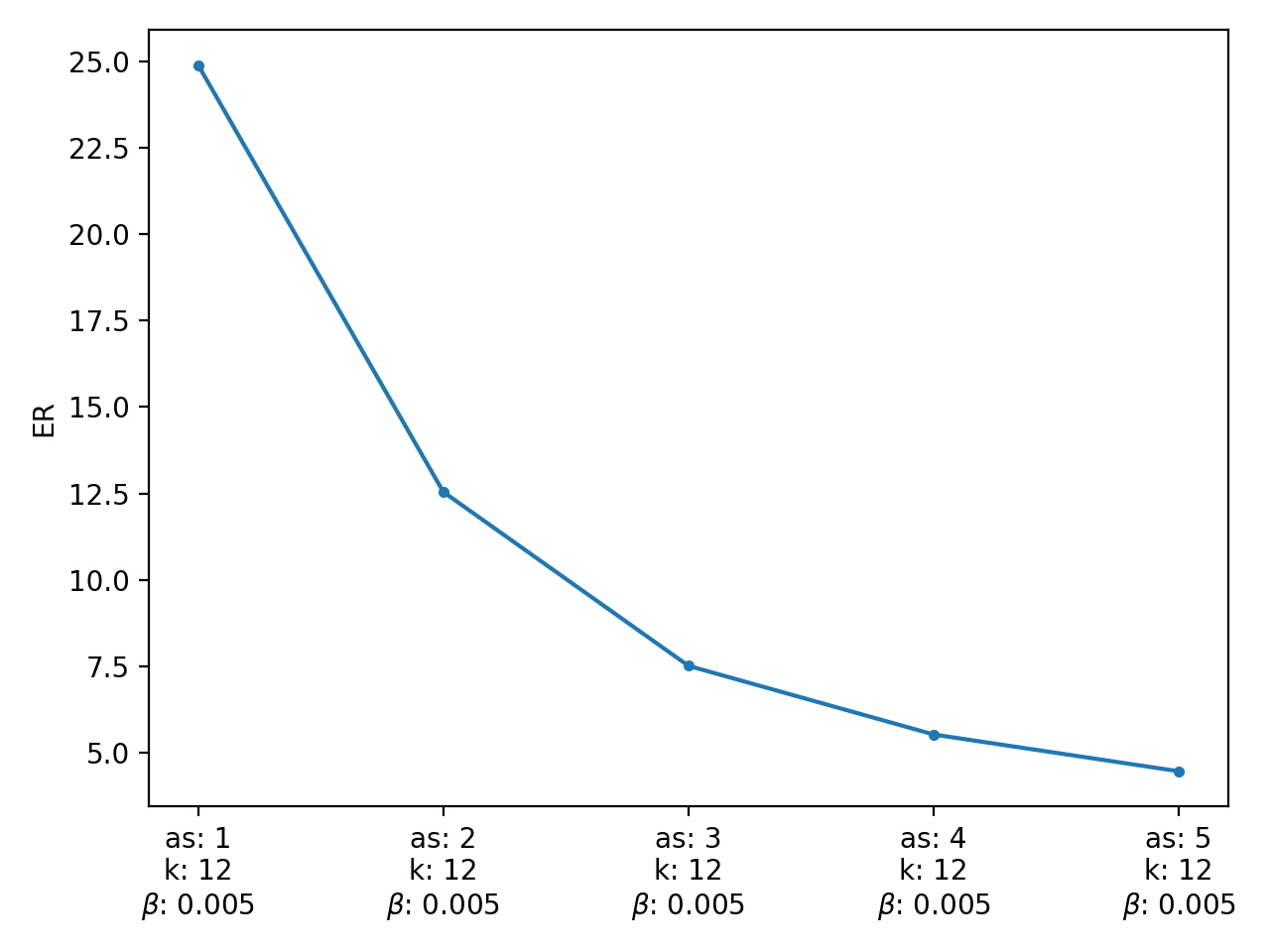}  \label{fig:noise-0-rel-0-5}} 
	\subfloat[$ap= 0.7$
	]{\includegraphics[width=.45\linewidth]{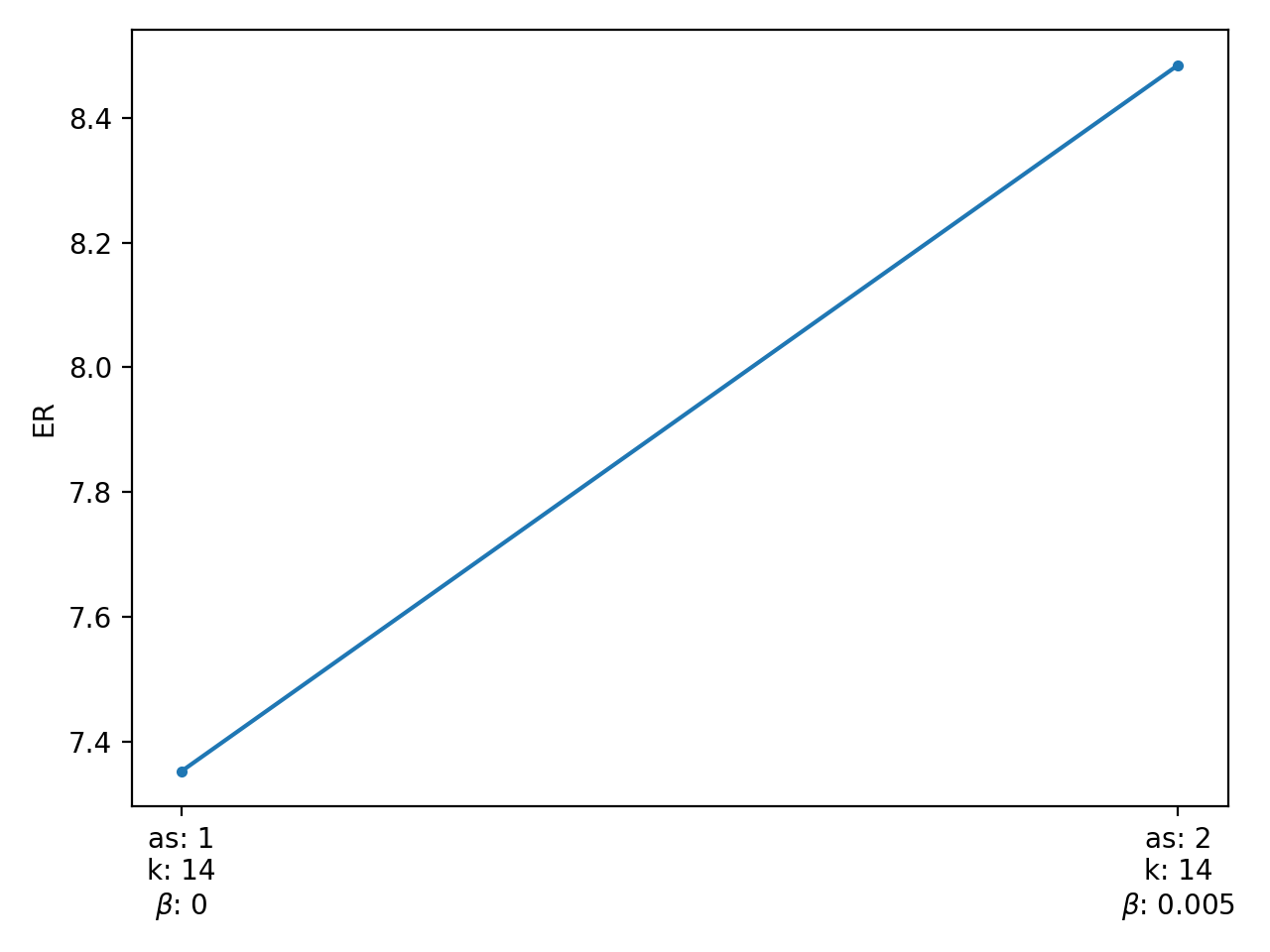}  \label{fig:noise-0-rel-0-7}} 
	\caption{
	Error ration for dataset with noise variance $\sigma_v^2=0$. 
	}
	\label{fig:results-noise-0}
\end{figure*}

\begin{figure}[t]
	\begin{subfigure}{.5\textwidth}

	\captionsetup[subfigure]{labelformat=empty}
	\centering
	\subfloat[ $ap= 0.1$
	]{\includegraphics[width=.45\linewidth]{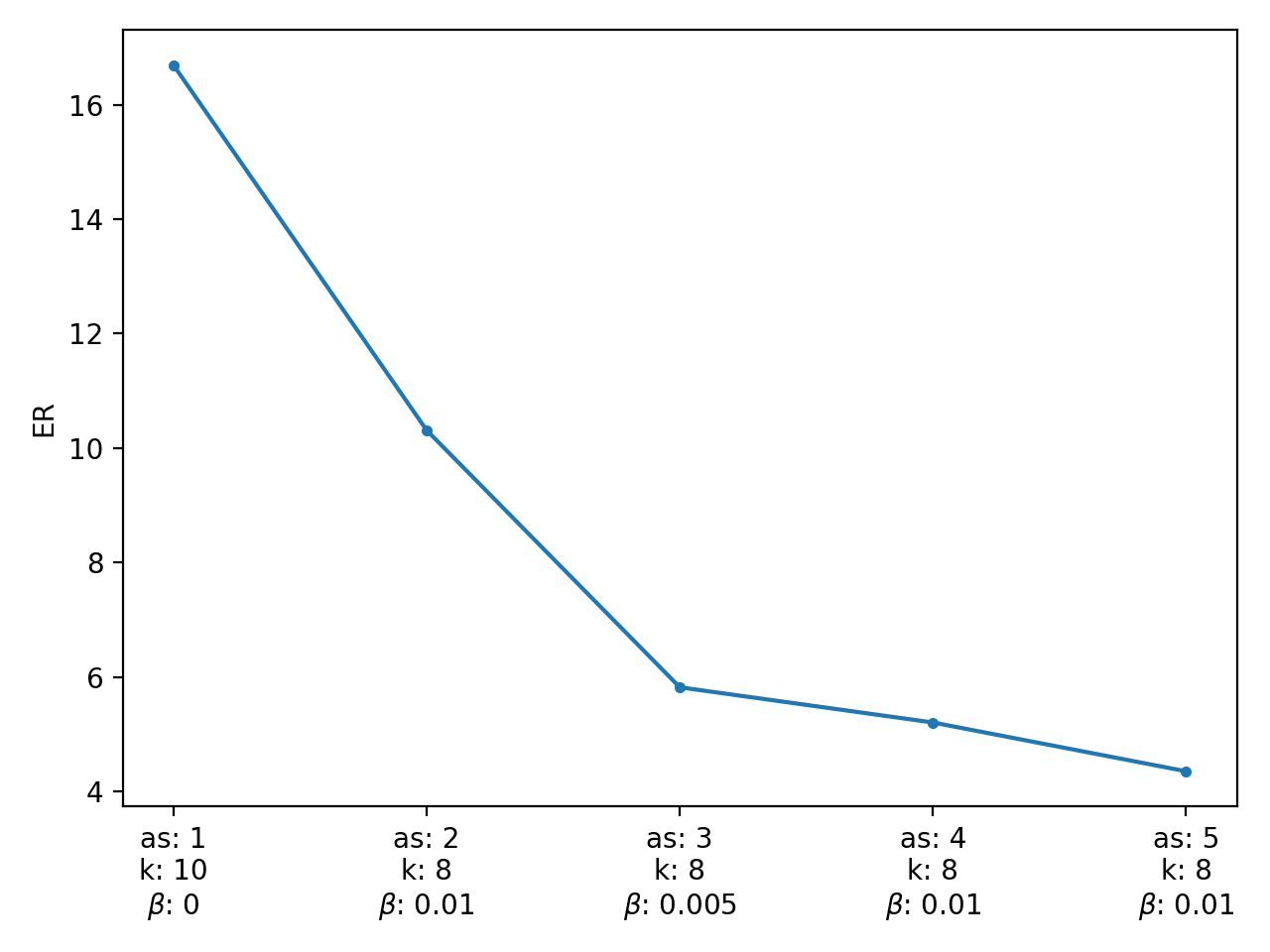}}
	\subfloat[ $ap= 0.5$
	]{\includegraphics[width=.45\linewidth]{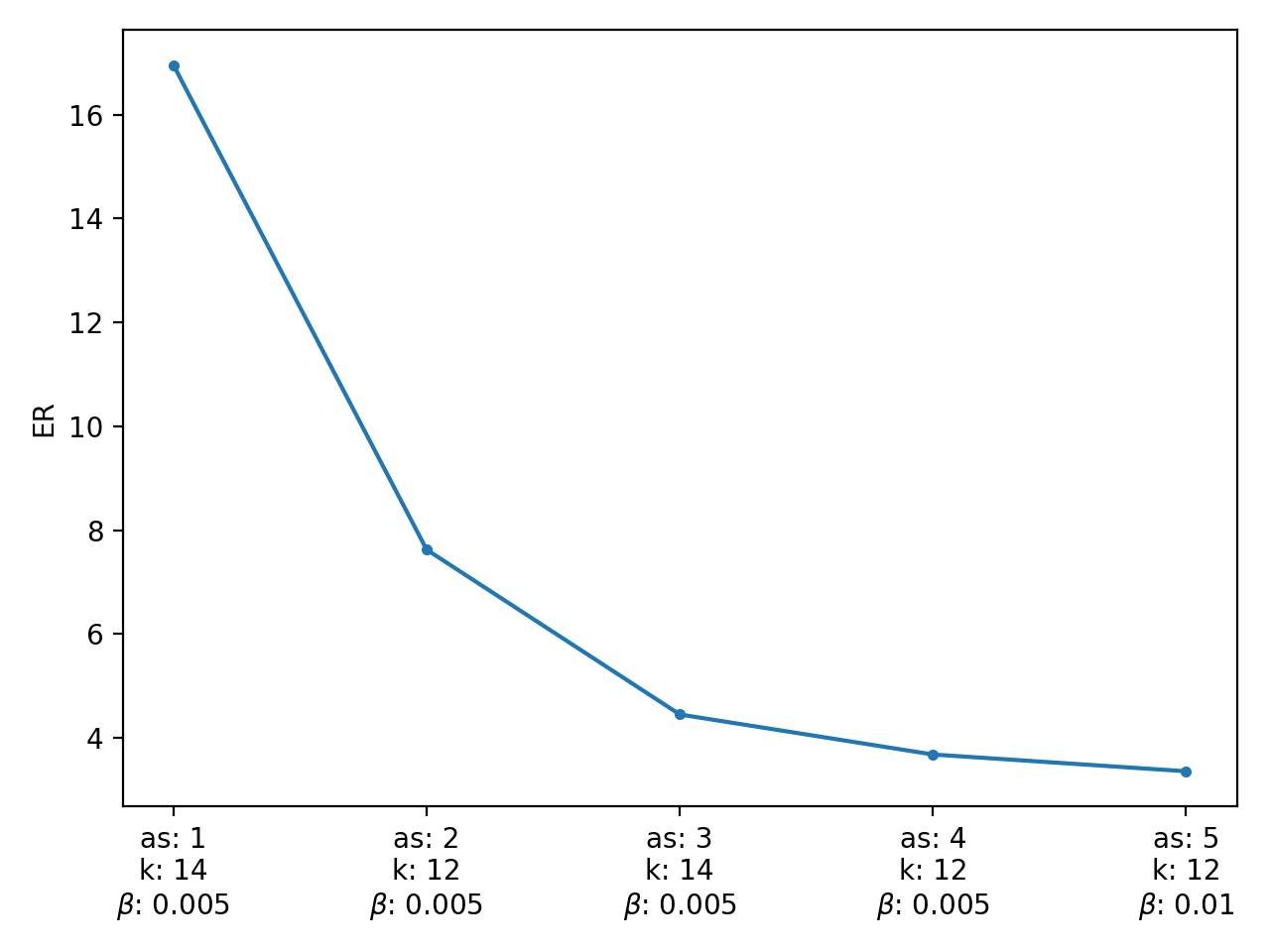}} \\
	\subfloat[ $ap= 0.3$
	]{\includegraphics[width=.45\linewidth]{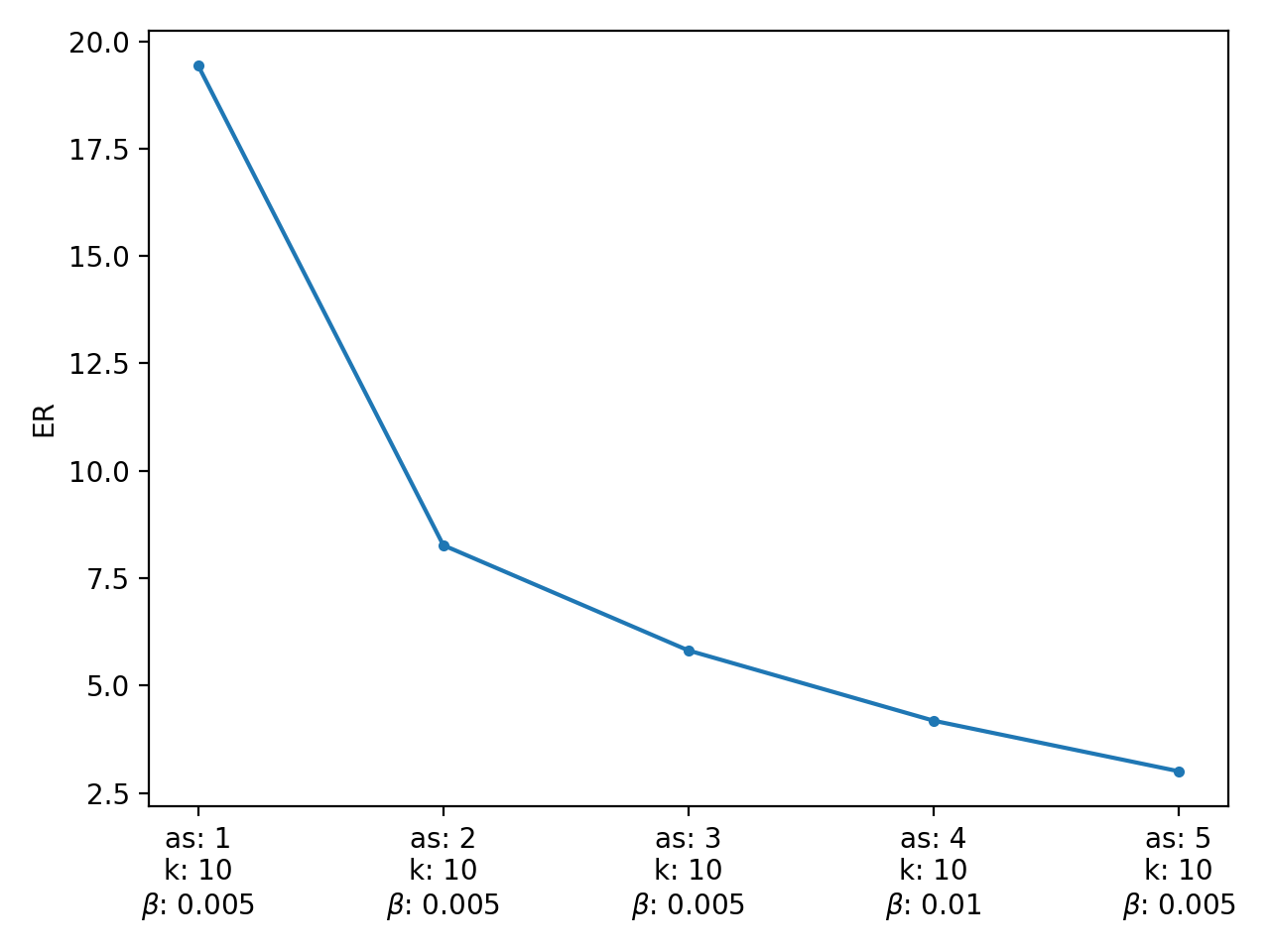}}
	\subfloat[$ap= 0.7$
	]{\includegraphics*[width=.45\linewidth]{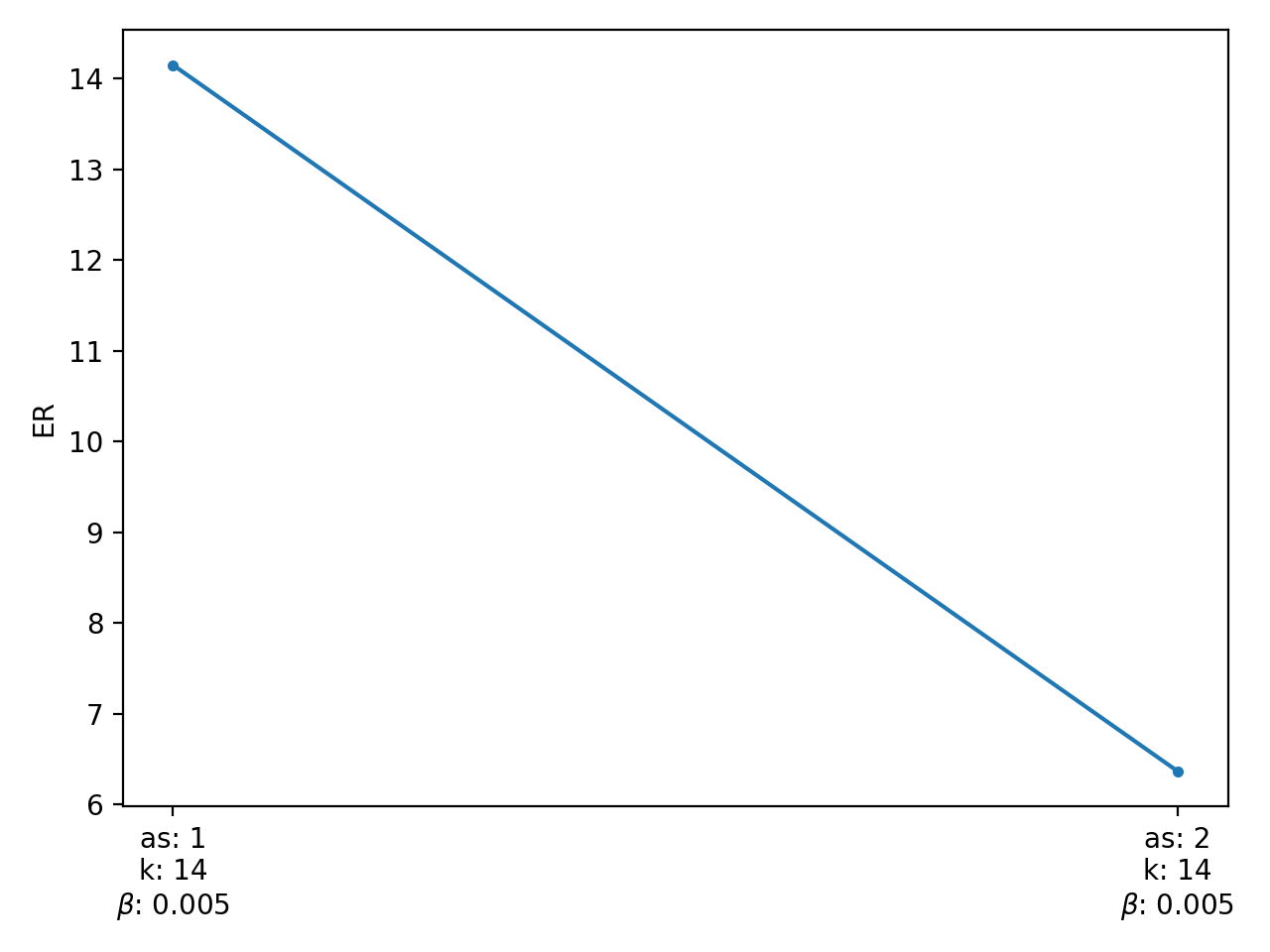}}
	\setcounter{subfigure}{0}
	\caption{
		$\sigma_v^2=0.1$
	}
	\label{fig:results-noise-0.1}
\end{subfigure}
\begin{subfigure}{.5\textwidth}
	\captionsetup[subfigure]{labelformat=empty}
	\centering
	\subfloat[ $ap= 0.1$
	]{\includegraphics[width=.45\linewidth]{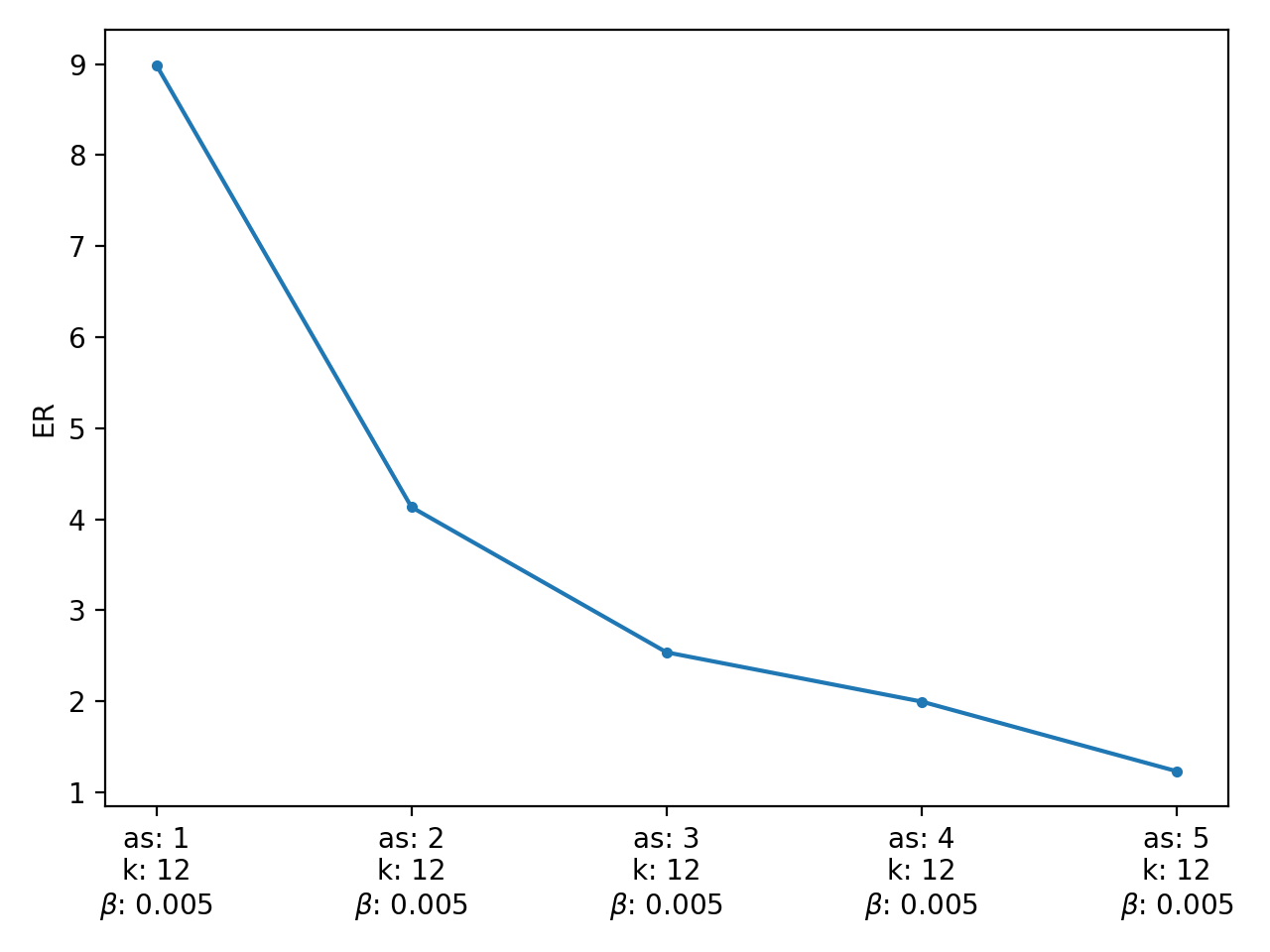}  \label{fig:noise-0-2-rel-0-1}} 
	\subfloat[ $ap= 0.3$
	]{\includegraphics[width=.45\linewidth]{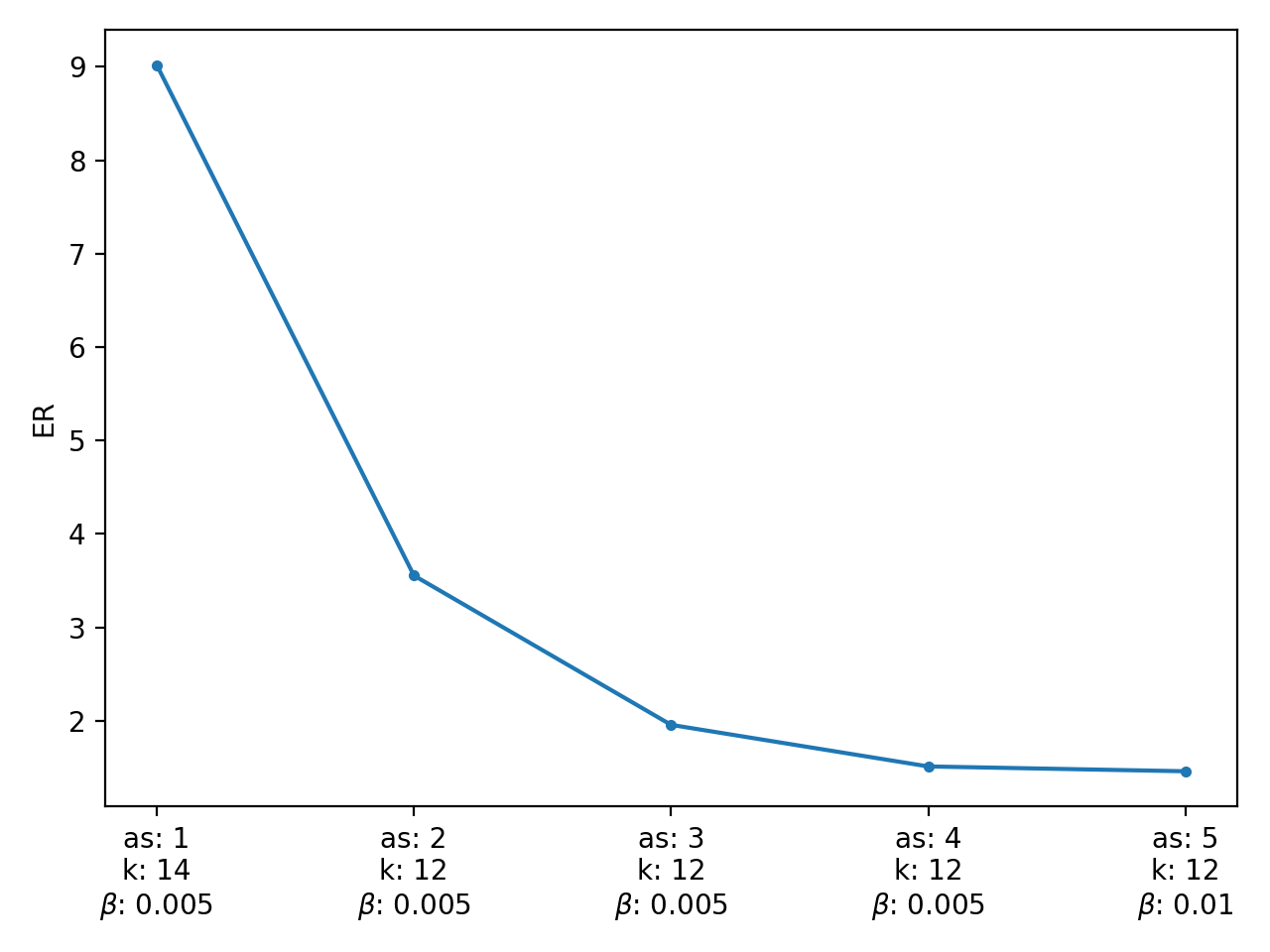}  \label{fig:Fnoise-0-2-rel-0-3}} \\
	\subfloat[ $ap= 0.5$
	]{\includegraphics[width=.45\linewidth]{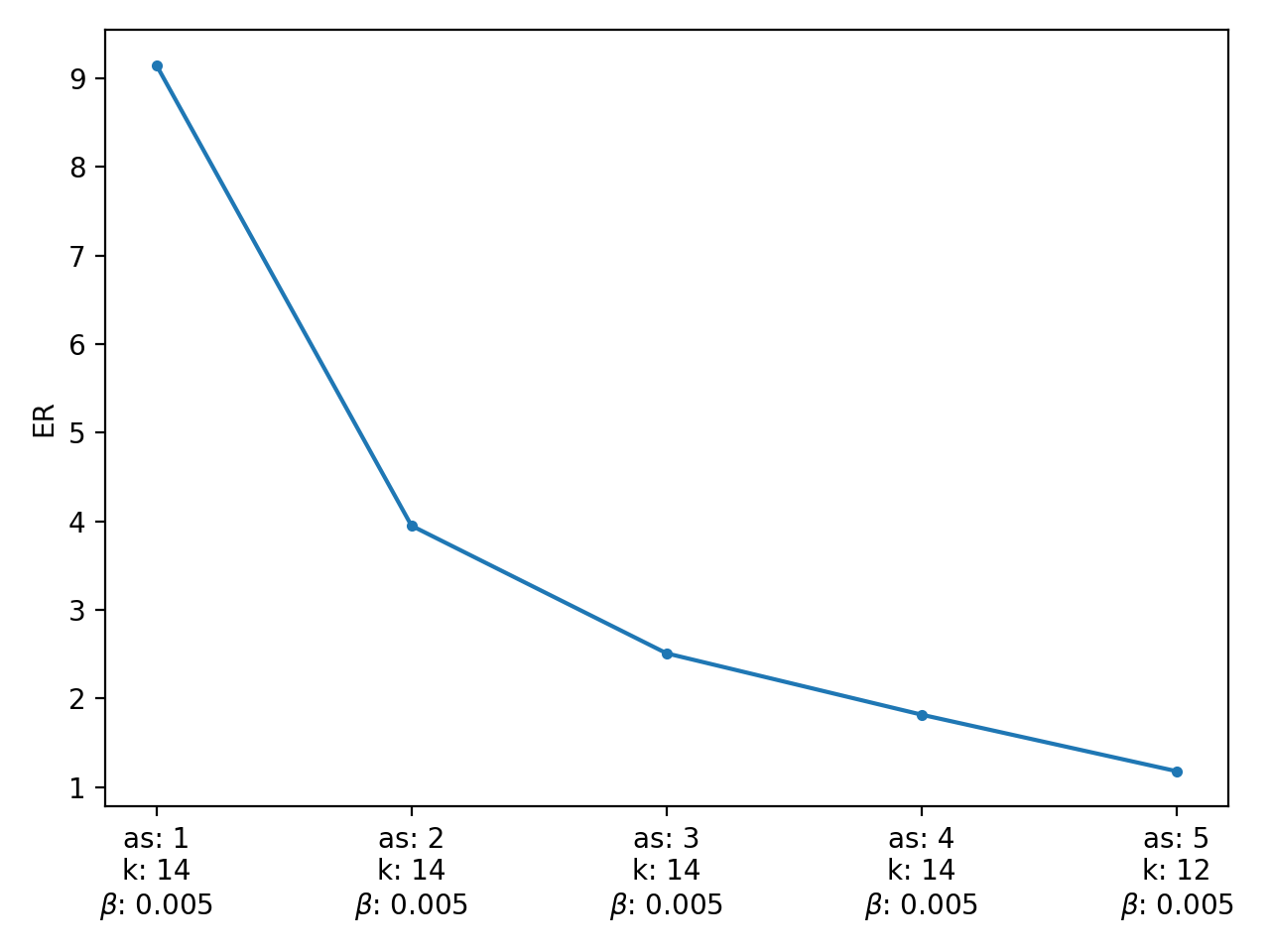}  \label{fig:noise-0-2-rel-0-5}} 
	\subfloat[$ap= 0.7$
	]{\includegraphics[width=.45\linewidth]{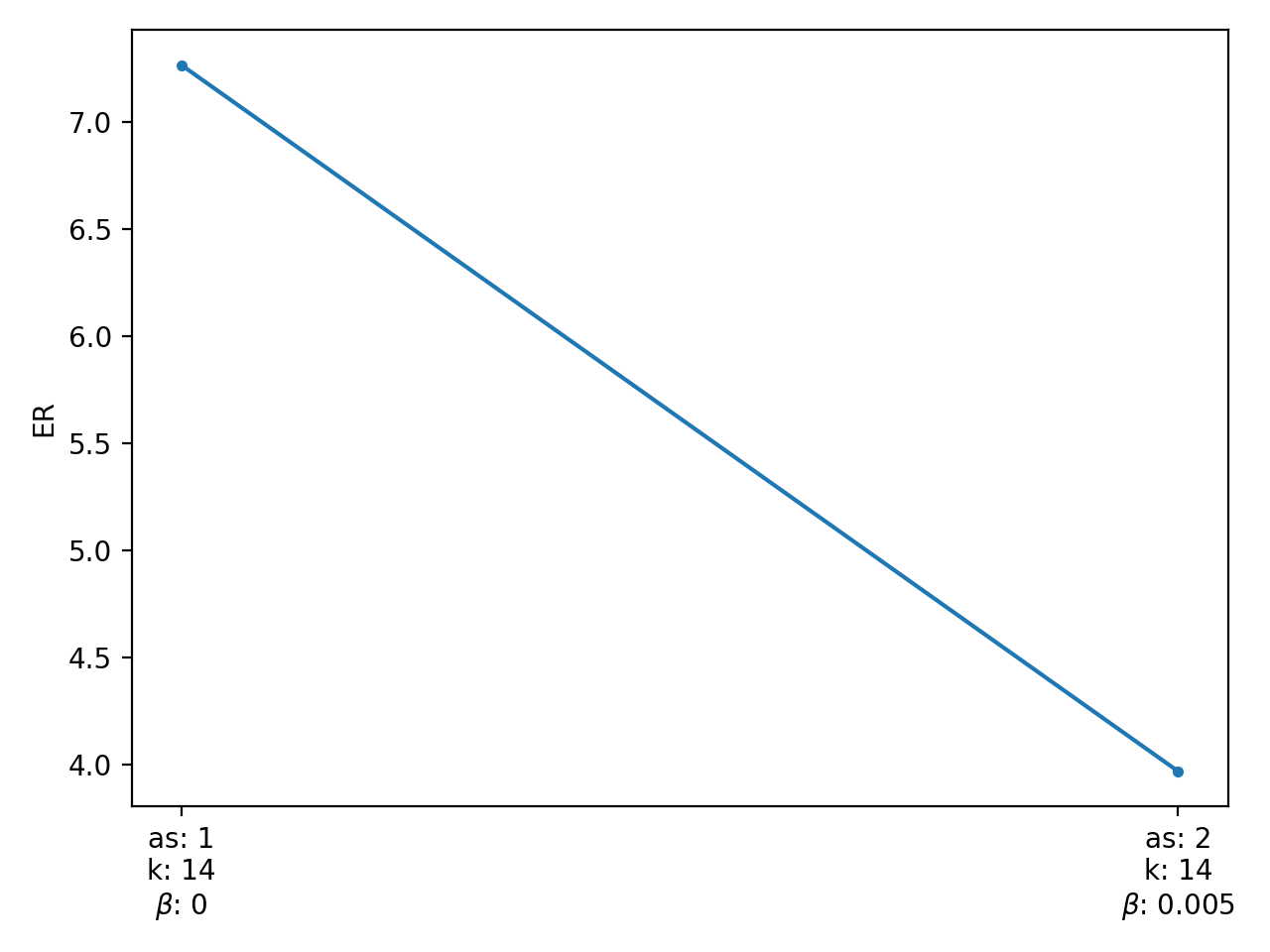}  \label{noise-0-2-rel-0-7}} 
	\setcounter{subfigure}{1}
	\caption{
		$\sigma_v^2=0.2$
	}
	\label{fig:results-noise-0.2}
\end{subfigure}
\begin{subfigure}{.5\textwidth}
	\captionsetup[subfigure]{labelformat=empty}
	\centering
	\subfloat[ $ap= 0.1$
	]{\includegraphics[width=.45\linewidth]{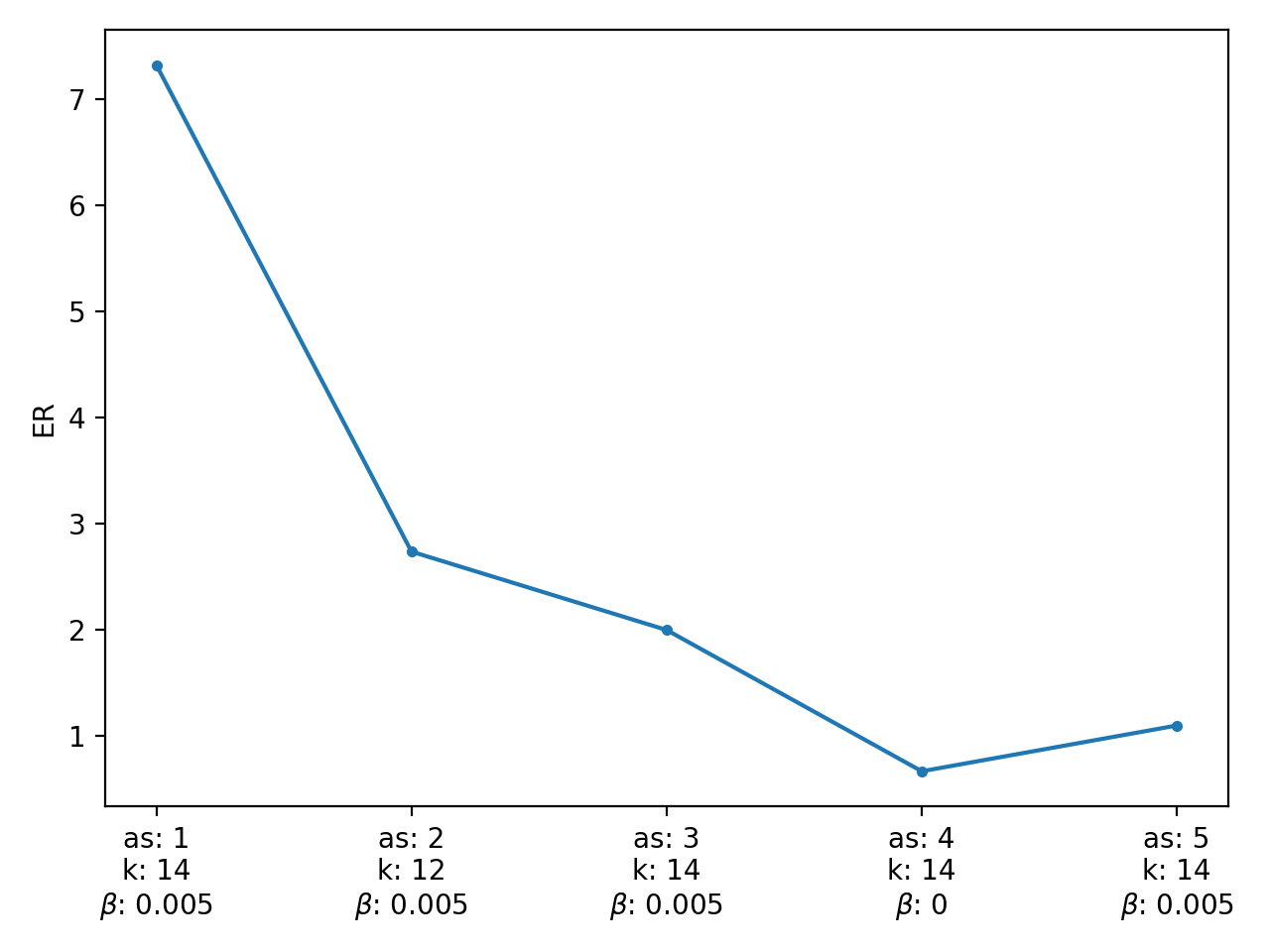}  \label{fig:noise-0-3-rel-0-1}} 
	\subfloat[ $ap= 0.3$
	]{\includegraphics[width=.45\linewidth]{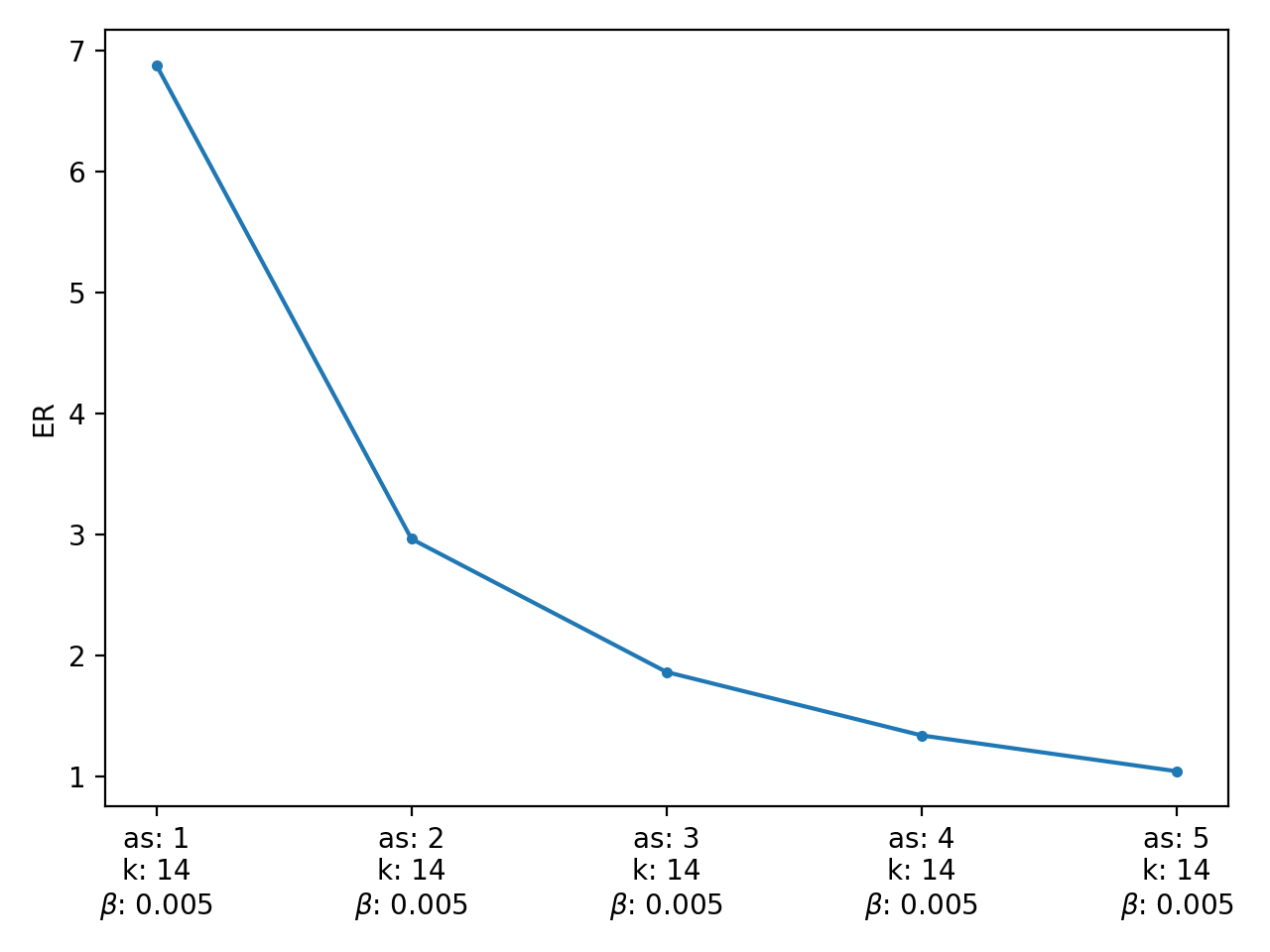}  \label{fig:Fnoise-0-3-rel-0-3}} \\
	\subfloat[ $ap= 0.5$
	]{\includegraphics[width=.45\linewidth]{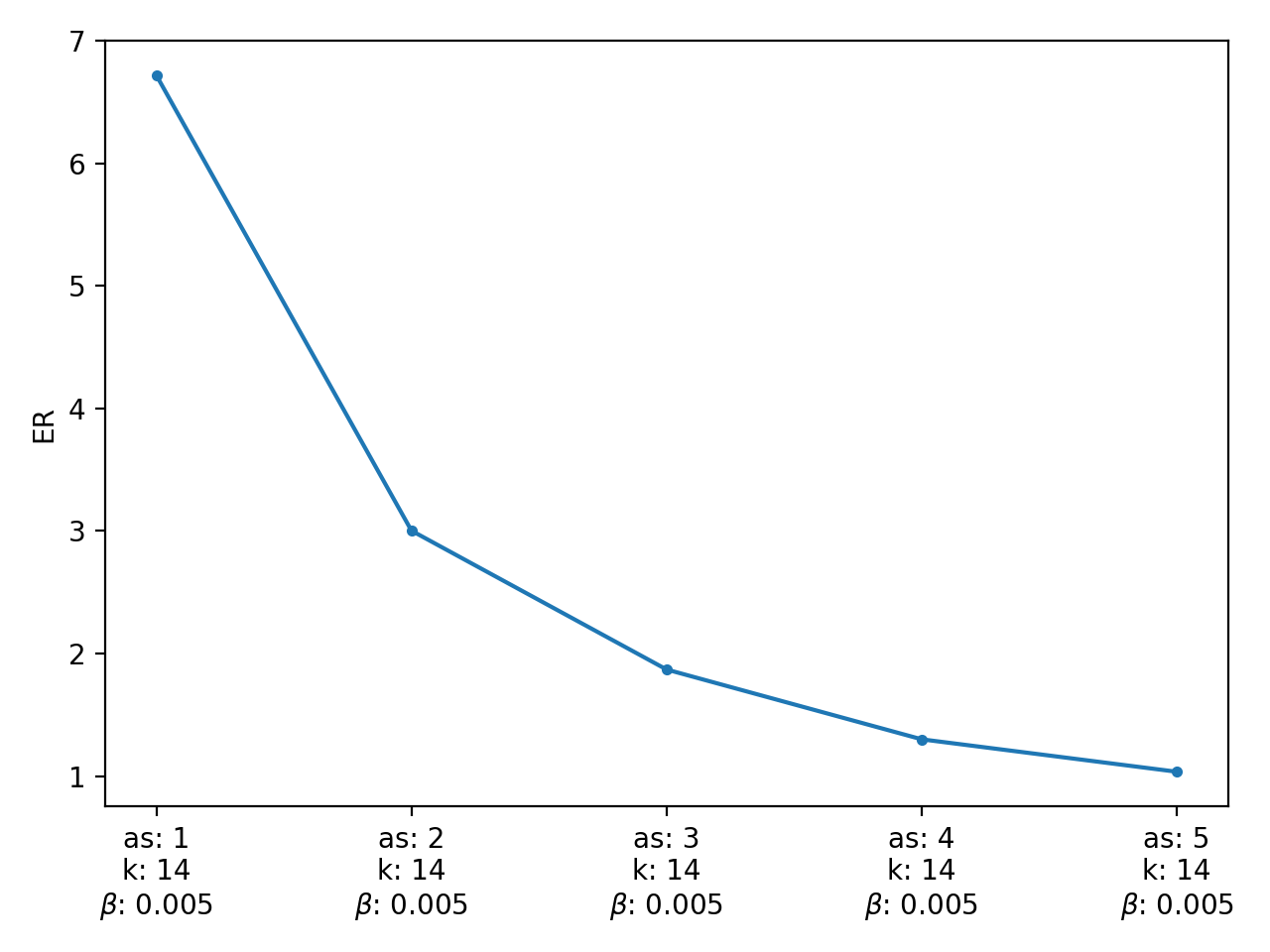}  \label{fig:noise-0-3-rel-0-5}} 
	\subfloat[$ap= 0.7$
	]{\includegraphics[width=.45\linewidth]{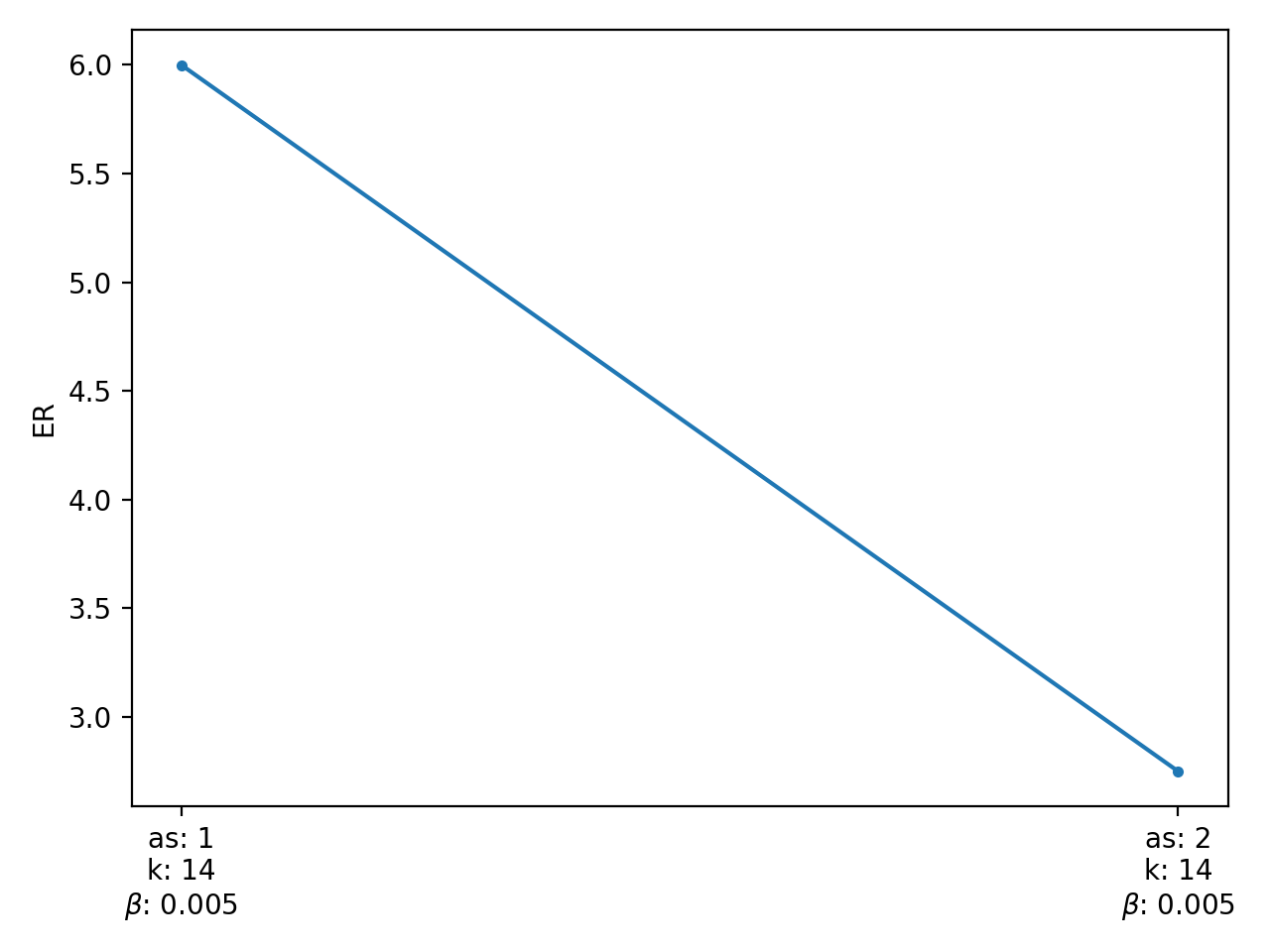}  \label{noise-0-rel-0-7}} 
	\setcounter{subfigure}{2}
	\caption{
		$\sigma_v^2=0.3$
	}
	\label{fig:results-noise-0.3}
\end{subfigure}
\begin{subfigure}{.5\textwidth}
	\captionsetup[subfigure]{labelformat=empty}
	\centering
	\subfloat[ $ap= 0.1$
	]{\includegraphics[width=.45\linewidth]{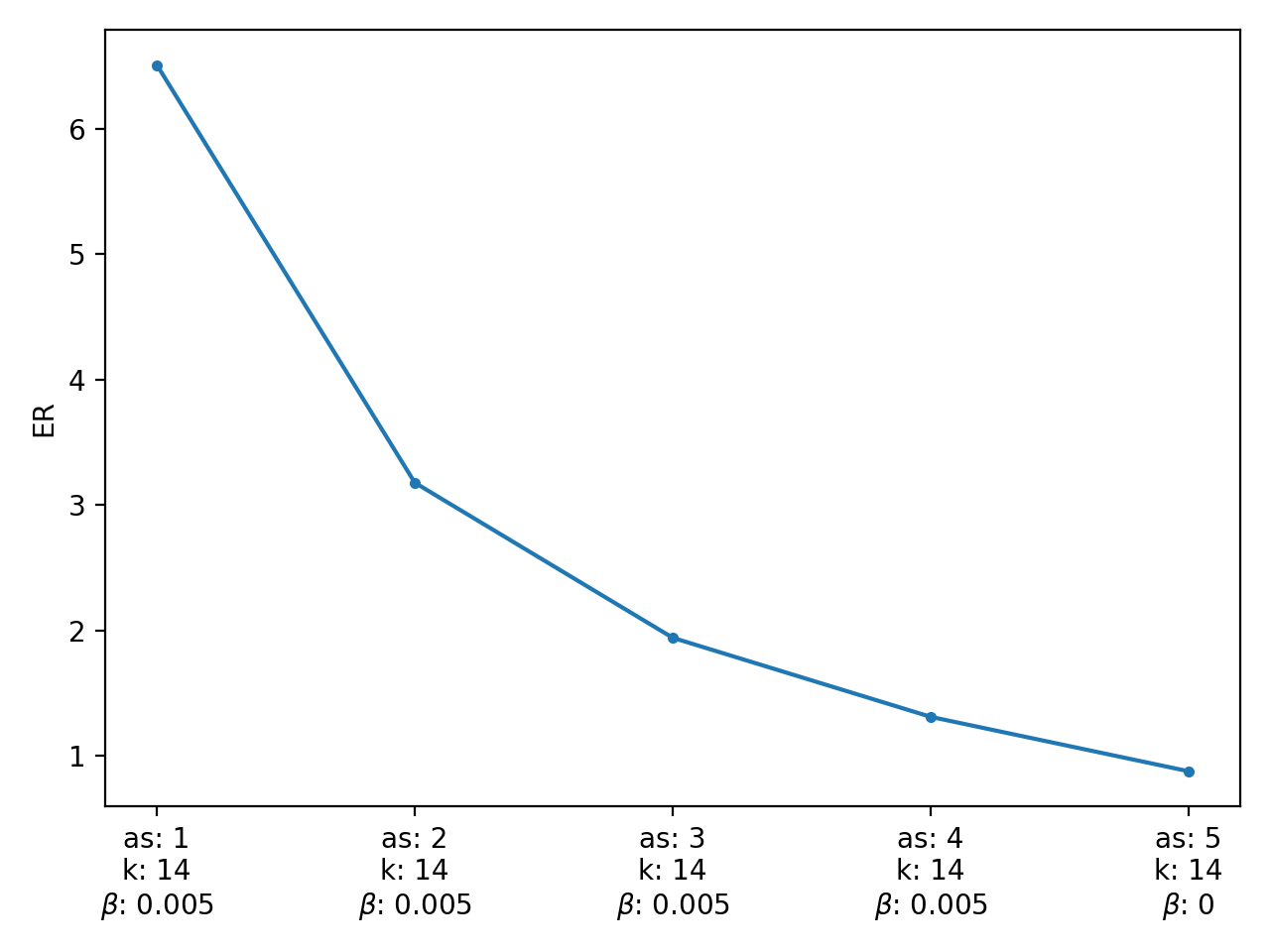}  \label{fig:noise-0-4-rel-0-1}} 
	\subfloat[ $ap= 0.3$
	]{\includegraphics[width=.45\linewidth]{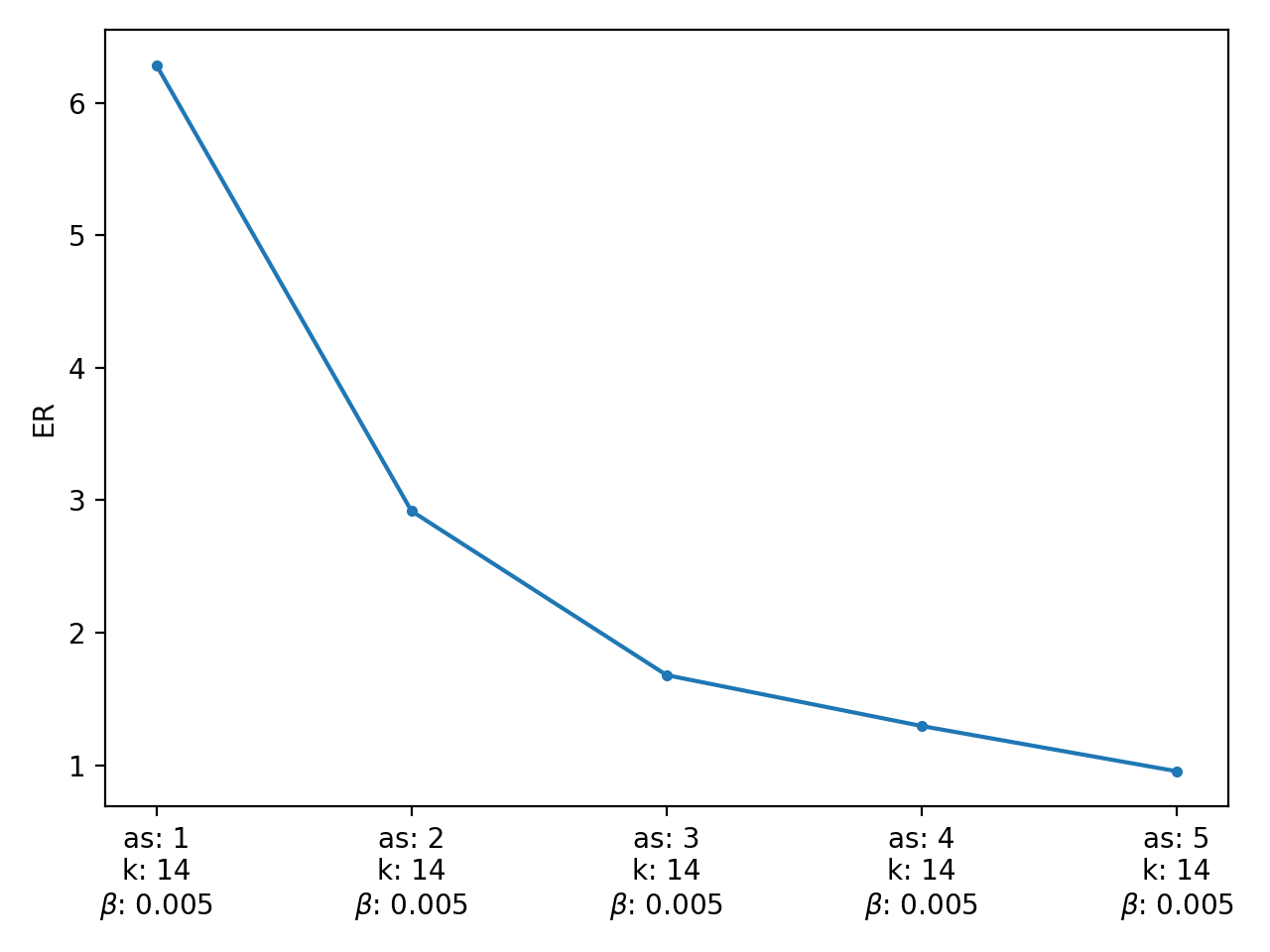}  \label{fig:Fnoise-0-4-rel-0-3}} \\
	\subfloat[ $ap= 0.5$
	]{\includegraphics[width=.45\linewidth]{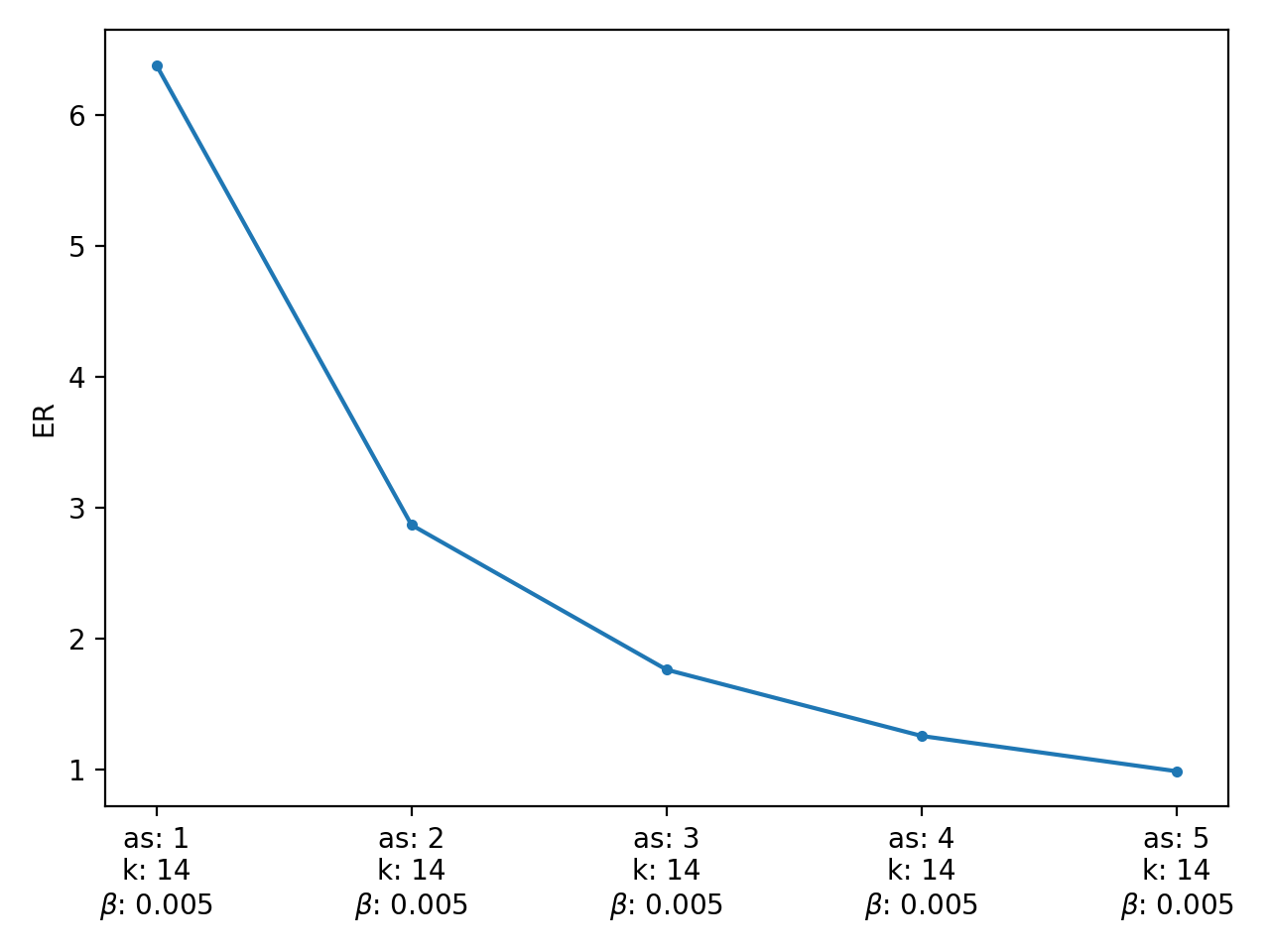}  \label{fig:noise-0-4-rel-0-5}} 
	\subfloat[$ap= 0.7$
	]{\includegraphics[width=.45\linewidth]{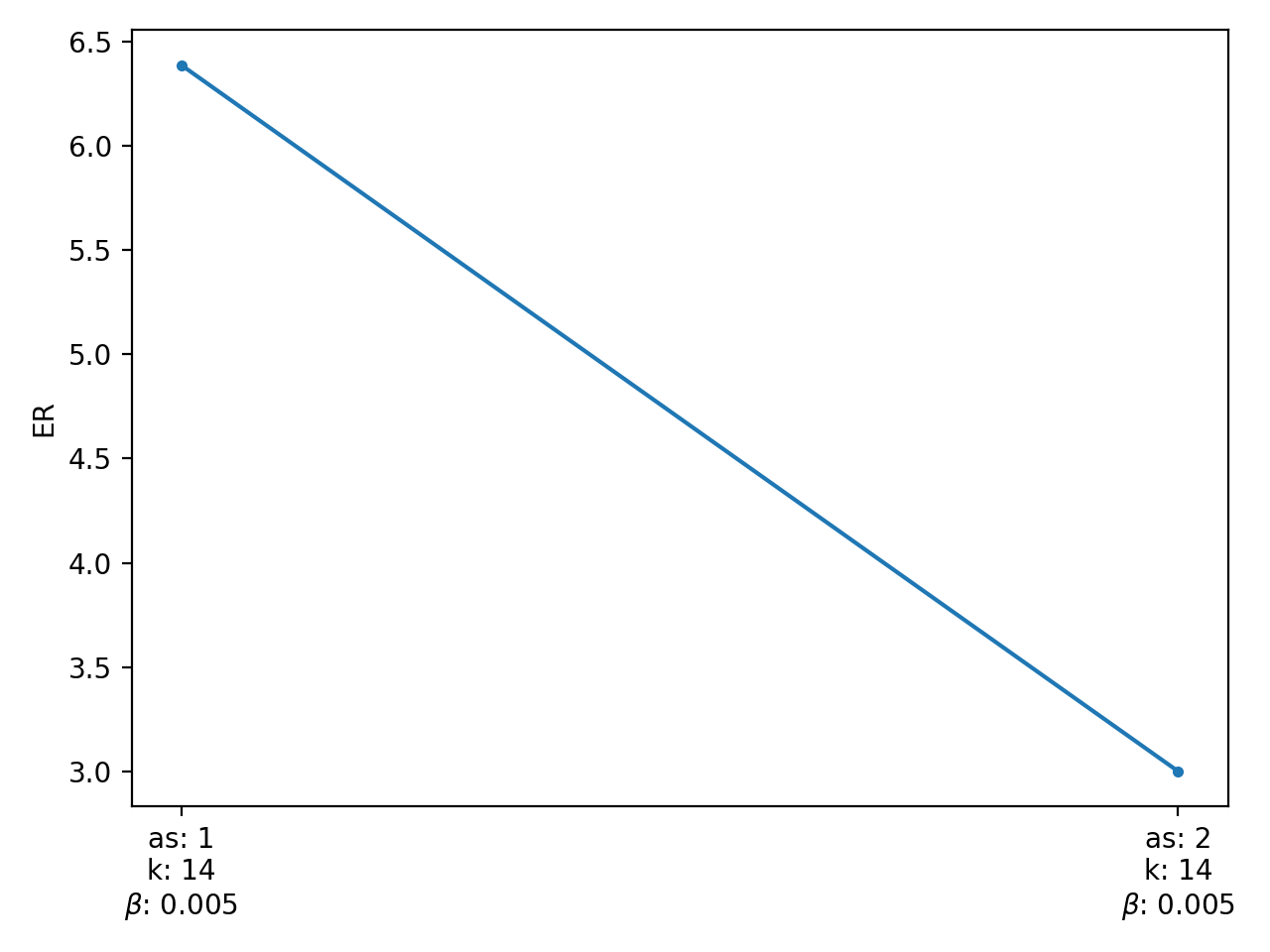}  \label{noise-0-4-rel-0-7}} 
	\setcounter{subfigure}{3}
	\caption{
		$\sigma_v^2=0.4$
	}
	\label{fig:results-noise-0.4}
\end{subfigure}
	\caption{
	Error ration for dataset with noise variance $\sigma_v^2 \in \{0.1,0.2,0.3,0.4\}$. 
	}
	\label{fig:results-noise-0.1-0.4}
\end{figure}

Since synthesized data is used for evaluation, we try to evaluate the SBSVD algorithm in different possible environments by using a range of different values as the network parameters. The results of the experiment are presented in figures \ref{fig:results-noise-0} and \ref{fig:results-noise-0.1-0.4} . Each diagram represents the resulting ER for a dataset with different network parameters noise variance $\sigma_v^2$, adversary proportion $ap$, and adversary stake $as$ (as the horizontal axis). In order to calculate ER, as explained in section \ref{ssec:algorithm-parameters}, for each set network parameters, we have iterated over the algorithm parameters $k \in \{6,8,10,12,14\}$, and $\beta \in \{0,0.005,0.01,0.02\}$. The resulting ER for the iteration with the least RSBE is depicted in the diagrams. Furthermore, the parameters for the optimal run are added to the horizontal axis alongside $as$. Notice that we assume at least $50\%$ of the staked tokens belong to honest users, and as a result, when $ap = 0.7$, then $as$ can be two at most.

First, we analyze the diagrams for the datasets with parameter $\sigma_v^2=0$ depicted in figure \ref{fig:results-noise-0}. In all of the diagrams, ER is greater than 1, indicating that the approximation error for honest users is lower than the adversary users. In fact, the values of RE are quite high, ranging from 25 in the best case to and 5 in the worst case. The dominant pattern is that with the increase in the value of $as$, RE is decreased. This is because $as$ determines the weight of adversary users, and as it increased, the approximation error for them gets closer to honest users. There are two exceptions in this figure : (1) in  diagram \ref{fig:noise-0-rel-0-3} in transition from $as=4$ to $as=5$, and (2) in the diagram \ref{fig:noise-0-rel-0-7} in transition from $as=1$ to $as=2$. In these transitions, the value of $\beta$ increases, too, which has caused the model to punish overfitting more severely. Adversary users benefit the most from overfitting, and when it is punished, their error increases, leading to a higher RE.

Furthermore, as the proportion of adversaries increases ($ap$), higher values of $k$ are used to achieve the optimal results. The reason for this is that when a larger proportion of users are adversary, a simple model with lower values of $k$ can not fit the users properly, and approximation error will be the most significant part of the formula \eqref{eq:RSBE}. Therefore, to keep RSBE low, the model goes towards overfitting by utilizing higher values of $k$. As a result, the approximation error is decreased for both honest and adversary users, but the degree of reduction is much higher for adversary users, which leads to the lower RE.

Figure \ref{fig:results-noise-0.1-0.4} shows the results for dataset with $\sigma_v^2 \in \{0.1,0.2,0.3,0.4\}$. As noise variance $\sigma_v^2$ increases $k$ increases too. The reason is that high values of noise fade the difference between honest and adversary users, and all of the votes seem to be cast randomly, and as a result, a simple model, with lower $k$ will yield a very high value of RSBE. Similar to the case of $ap$, the model will opt higher values for $k$ to decrease RSBE at the cost of a more complex model. In the dataset with $\sigma_v^2=0.4$ (figure \ref{fig:results-noise-0.4}) the behavior of honest and adversary user are quite the same, as RE is close to 1 when all of the users stake five tokens. In an environment with this property, the amount if stakes by honest and adversary users are the most determinant factor in the performance of the algorithm. As explained in section \ref{sssec:staking}, stakes mechanism disincentives adversaries from staking a high amount of tokens, and as a result, honest users are expected to have a higher amount of stakes compared to adversaries. Therefore, we expect the model to have a decent performance even if the honest and adversary behavior have inconspicuous borders.

 \section{Related Works} \label{sec:related-works}

The literature pertaining to our work includes three main areas: (1), blockchain-powered news networks; (2), layer-2 blockchain scalability solutions; and (3), detecting fraudulent users in review networks.

\subsection{News Networks on Blockchain}

Bhatia et al. \cite{Sapien} introduce Sapien, which is a social news platform based on the Ethereum platform. In this network, users gain reputations through their activities. These reputations will then play a crucial role in the process of detecting fake news, rewarding users, and community organization. Mondrus et al. \cite{trive} describe an environment called trive in which users ask for verification of news items by submitting some tokens. The news item goes through a verification game that provides incentives for research into the item and involves several actors. If the process finishes without any conflict between the actors, everyone is rewarded, but if the actors run into a conflict, a randomly selected pool of actors decide who is just, based on the majority vote. Singh et al. \cite{DNN} propose a distributed news network (DNN) in which seven randomly isolated reviewers decide on whether to accept a news item or reject it based on the item's compliance to a set of content guidelines. The majority is the deciding factor in this process. In Media Sifter network proposed by Camus \cite{MediaSifter}, users put bounties on news items and incentivize investigators to provide evidence for the item. The pieces of evidence will, in turn, be judged by a series of randomly selected reviewers, and users can access them. Steem network \cite{steemit}, despite not being a mere news network, is one of the first blockchain-based social networks which used the notation of Steem Power as a long-term investment that could determine the extent of user's effect in the network.

Our work is different from these works in the sense that they are all susceptible to the dominance of united users meaning a large enough group of users can determine what content should be promoted in the network and what should be undermined. Besides, Trive, DNN, and Media MediaSifter require news items to be reviewed one by one, which can stymie the scalability of the network.

\subsection{Layer-2 Scalability Solutions}

In \cite{poon2016bitcoin}, Poon and Dryja introduce Lightning Network, a micropayment channel for Bitcoin. Users should create a between themselves and lock some Bitcoins. After the channel creation, users can have trustless instant off-chain payments, which are enforced on the blockchain through the channel creation process. Lightning Network was designed with off-chain payments in mind, and it is unable to provide scalability for applications with more complex logic. To address this problem, Poon and Buterin proposed Plasma \cite{poon2017plasma}. Plasma is a series of smart contracts that enable having several blockchains on top of the root blockchain. Plasma takes advantage of a side-chain that is managed through a proof-of-authority or proof-of-stake consensus mechanism. The hashes of the blocks created in the side-chain are committed to the root blockchain. To have a plasma application, the rules of the application should be encoded into a smart contract. In the case of misbehavior, from managers or users, a fraud proof can be generated, which proves the behavior was against the network rules using the committed block hashes. However, encoding the rules of an application in a way that they are enforceable using only block hashes can introduce severe challenges.

In \cite{teutsch2018truebit} Teutsch and Reitwießner introduce Truebit as a solution to verify off-chain computations on blockchain. Using this solution, the solver submits the result of his computation to the blockhain. After the submission, a series of challenges verify the validity of the computations and start a verification game if the results are invalid. The result of the game is finding the line with a conflicting result in their computations, and this line is executed on blockchain to find out who is just.

In \cite{ZKROllUp} Buterin proposes a method to achieve approximately 500 transactions per second on the Ethereum network. The idea is that instead of using a signature for each transaction, ZK-SNARKs can be used to mass validate transactions by generating a single proof that enforces the validity of all of the transactions.

\subsection{Detecting Fraudulent Users in Review Networks}

One of the approaches toward this problem is using supervised machine-learning methods \cite{siering2016detecting,li2011learning,lin2014towards,ho2016computer,lau2011text}. The problem in the way of adopting these methods in our networks lies in the fact that they need labeled data in order to train their model. In a blockchain-based setting, reaching consensus on what data to use for training could introduce several new problems and lead to unfair models.

Wu et al. \cite{wu2010distortion} tried to detect suspicious hotel reviews by finding the reviews that have a more significant role in the model's distortion. Akoglu et al. \cite{akoglu2013opinion} use a probabilistic graphical model based on network effects to find fraudulent users and spam reviews in online review networks. Wang \cite{wang2011review} quantifies the trustiness of reviewers, the honesty of reviews, and the reliability of stores in a store review network, and exploits an iterative computation framework to calculate these values.a

Wu et al. \cite{wu2010distortion} tried to detect suspicious hotel reviews by finding reviews that have a more significant role in the model's distortion. Akoglu et al. \cite{akoglu2013opinion} use a probabilistic graphical model based on network effects to find fraudulent users and spam reviews in online review networks. Wang \cite{wang2011review} quantifies the trustiness of reviewers, the honesty of reviews, and the reliability of stores in a store review network, and employs an iterative computation framework to calculate these values.

The final model in these approaches assigns a specific value to each item, and each user who deviates from these values is labeled as fraudulent. This viewpoint is not applicable to our network since we need to take different perspectives into account.

Another approach toward assessment of a user's behavior is using the notions of structural balance \cite{cartwright1956structural} and weak structural balance \cite{davis1967clustering}. The idea is rooted in a theory in psychology \cite{heider1946attitudes} and claims that signed graphs in social interactions tend to have properties that enable us to separate the nodes into two (or more in case of weak structural balance) disjoint sets so that all of the edges between nodes from different sets are negative, and all of the edges between nodes in the same set are positive. There are methods to measure the structural balance of a graph \cite{facchetti2011computing, ma2015memetic, sun2014fast}, and at first glance, it seems that these methods can be used to indicate fraudulent users by computing the structural balance for each user. But, there are two problems with this approach regarding our network. First, our graph is extremely spare, which leads to the poor performance of these methods. Second, since a news item has multiple aspects, it is not plausible to expect structural balance in the proposed news graph.

\section{Threats to Validity} \label{sec:threats to validity}

In this section, we analyze the threats to the validity of our experiment.

\subsubsection{Construct Validity}
the construct validity is the concern about the suitability of the evaluation measures. The objective of the experiment is to asses the model's ability to discern honest users from adversary users in a graph-based social news aggregator. Because of the novelty of this study, no prior study has been carried out with this objective, and we introduced a new measure called ER (error ratio) to evaluate the results. This measure determines the ratio of the RMSE of the adversary users to the RMSE of the honest users, which is in accord with the objective of the experiment, and the results corroborate the theoretical expectation.

\subsubsection{Internal Validity}
The internal validity is the concern about whether the changes in the dependent variable were because of the changes made to the dependent variable. The data used in the experiment are synthesized through a controlled process, and variables are assigned according to the experiment objective. Honest and adversary users are selected from an identical distribution, and only the changes in independent variables (proportion of adversary users, adversary users' stake, and adversary users' voting behavior) have rendered the experiment results.

\subsubsection{External Validity}

The external validity is the concern about the generalization of the results. Since there was no real-world network with our desired properties, we were forced to synthesize the data for our evaluation, which can impose a threat to the external validity of our study. Nevertheless, to mitigate the threat, we employed several techniques to enhance the generalizability of our results.

In the process of data generation, we utilized tools and suppositions to bring the synthesize data as close as possible to real-world data. With this in mind, the news graph was generated using RTG \cite{akoglu2009rtg}, which generates realistic graphs analogous to real-world graphs. In addition, with respect to the remarkable performance of matrix factorization on a range of different review networks, the votes matrix was generated as a noisy low-rank matrix. Furthermore, by changing the proportion of adversary users, their stakes, and the variance of the noise added to the votes matrix, the performance of our proposed algorithm was analyzed on a wide range of possible environments that the networks might operate in.

\section{Future Work} \label{sec:future-work}
In this study, we tried to propose a novel type of social news aggregator. Several enhancements can be applied in future work:

\begin{itemize}
\item In this study, we considered random voting as the only detrimental behavior that can be carried out by adversary users. More study is needed for detecting and analyzing other behaviors that could damage the network.

\item We assigned a fixed value to the network parameter $\lambda$. The design of a mechanism for updating this parameter as the network evolves is a future work to pursue.

\item One of the cons of the proposed network is that all of the votes are submitted directly to the main blockchain network, which can be costly. Embedding an integration mechanism like zk-rollup or using plasma in the network can help to address this problem.

\item A representative large-scale case study needs to be carried out to analyze the incentive mechanism on real-world data.

\end{itemize}

\section{Conclusion} \label{sec:conclusion}

In this work, we introduced a graph-based social news aggregator to address several problems of the current social news aggregators. News items are the nodes of the graph, and users vote on the edges, which represents the relation between news items. This approach mitigates the effect of the user's bias on news items' importance in the network. In addition, we use blockchain as the management layer to bring decentralization and transparency to the network and lay the foundation of the incentive mechanism by enabling cryptoeconomics. Next, we introduced our incentive mechanism, which employs a matrix factorization model to asses the behavior of users, and a layer-2 fraud proof protocol, which permits on-chain verification for matrix factorization computations which are done off-chain. The incentive mechanism, along with the layer-2 fraud proof protocol, incentivizes users to act honestly to gain more financial benefits. At last, we presented the time and space analysis of the incentive mechasim, and its performance on a range of synthesized datasets, which are generated to simulate different environments that the network might operate in.

\bibliographystyle{unsrt}  
\bibliography{references}  

\end{document}